\documentclass[prd,superscriptaddress,nofootinbib,a4paper,preprintnumbers]{revtex4} 

\usepackage[dvipdfmx]{graphicx}
\usepackage{graphicx}
\usepackage{dcolumn}
\usepackage{bm}
\usepackage{latexsym}
\usepackage{amsfonts}
\usepackage{amssymb}
\usepackage{amsmath}
\usepackage{bbold}
\usepackage{ulem}
\usepackage{xcolor}
\usepackage[utf8]{inputenc} 
\usepackage{xspace}

\newcommand{\A}{{\mathcal{A}}}
\newcommand{\ct}{{\tilde{c}}}
\newcommand{\CC}{{\mathcal{C}}}

\DeclareRobustCommand{\erase}{\bgroup\markoverwith{\textcolor{red}{\rule[.5ex]{2pt}{0.4pt}}}\ULon}

\allowdisplaybreaks[1]
\begin{document}

\title{Vainshtein Mechanism in Generalised Massive Gravity}

\author{A. Emir G\"umr\"uk\c{c}\"uo\u{g}lu}
\email{emir.gumrukcuoglu@port.ac.uk}
\affiliation{Institute of Cosmology and Gravitation, University of Portsmouth\\ Dennis Sciama
	Building, Portsmouth PO1 3FX, United Kingdom}

\author{Rampei Kimura}
\email{rampei@aoni.waseda.jp}
\affiliation{Waseda Institute for Advanced Study, Waseda University\\
	19th building, 	1-21-1 Nishiwaseda, Shinjuku-ku, Tokyo 169-0051, Japan
}

\author{Michael Kenna-Allison}
\email{michael.kenna-allison@port.ac.uk}
\affiliation{Institute of Cosmology and Gravitation, University of Portsmouth\\ Dennis Sciama
	Building, Portsmouth PO1 3FX, United Kingdom}

\author{Kazuya Koyama}
\email{kazuya.koyama@port.ac.uk}
\affiliation{Institute of Cosmology and Gravitation, University of Portsmouth\\ Dennis Sciama
	Building, Portsmouth PO1 3FX, United Kingdom}

\date{\today}

\begin{abstract}
We present a non-linear analysis of perturbations around cosmological solutions in Generalised Massive gravity. This Lorentz invariant theory is an extension of de Rham, Gabadadze, Tolley massive gravity that propagates $5$ degrees of freedom while allowing stable open FLRW cosmologies. 
For a minimal model that supports a self-accelerating background, we study the dynamics of non-linear perturbations. We find that the equation for the scalar graviton is distinct from the analogues in Horndeski and DHOST theories. We numerically solve the equation to find a new type of nonlinear solution for the scalar mode, and confirm the presence of a Vainshtein screening mechanism. We show that the PPN parameter approaches its GR value at solar system scales and satisfies the current bounds. 
\end{abstract}

\maketitle

\section{Introduction}

The cosmological constant is one of the pillars of concordance cosmology and provides the simplest model of the present-day accelerated expansion \cite{Perlmutter:1998np,Riess:1998cb}. Despite its phenomenological success, the cosmological constant is theoretically inadequate due to the excessive tuning necessary to match the observed value and its extreme sensitivity to the unknown high energy physics \cite{Martin:2012bt}.
An alternative approach to account for the acceleration is to introduce long distance modifications of gravity. By increasing the number of degrees of freedom with respect to General Relativity (GR), the equation of state of dark energy can be modified as well as the gravitational potentials and the evolution of density perturbations \cite{Koyama:2015vza}. A convenient framework to study all of these effects, which are largely constrained to the scalar sector, is the scalar-tensor theory class \cite{Langlois:2018dxi}. However, the recent discovery of the GW170817 gravitational wave event with an electromagnetic counterpart \cite{LIGOScientific:2017vwq} has ruled out a significant number of possible interactions between the metric tensor and the scalar field in these theories \cite{Creminelli:2017sry,Ezquiaga:2017ekz,Creminelli:2018xsv,Creminelli:2019nok,Creminelli:2019kjy}. The only remaining terms in this phenomenological framework are relatively simple extensions of a cosmological constant that cannot be falsified.

Alternatively, by spreading the modifications to the tensor sector, one can introduce unique interactions between different polarisations of the graviton that are not present in a scalar-tensor set-up and potentially evade some of the restrictions from observations. In this paper, we focus on such an example, the theory of massive spin--2 fields, which is arguably the simplest modification of GR at large distances. The original non-linear and Lorentz invariant massive gravity theory was introduced by de Rham, Gabadadze and Tolley (dRGT)  \cite{deRham:2010ik,deRham:2010kj}. The potential for the graviton is defined in reference to a fiducial metric $f_{\mu\nu}$, and depends on four St\"uckelberg scalars $\phi^a$ via
\begin{equation}
    f_{\mu \nu}\equiv \eta_{ab}\partial_{\mu}\phi^a\partial_{\nu}\phi^b\,, 
    \label{eq:f-defined} 
\end{equation}
where $a,b=0,1,2,3$ are the scalar field space indices and $\eta_{ab}$ represents the Minkowski metric. The Poincar\'e invariance in the field space is transferred to the physical metric $g$ via interactions constructed out of $g^{-1}f$. The dRGT potential involves a tuning that renders one of the degrees of freedom auxiliary, thus avoiding the Boulware-Deser instability that contaminated the early incarnations of massive gravity \cite{PhysRevD.6.3368}. The theory thus has $5$ propagating degrees of freedom around flat space-time and the agreement with the solar system tests is ensured by a non-linear Vainshtein screening \cite{Vainshtein} which turns off the scalar polarisation of graviton at scales where the graviton mass is negligible \cite{Babichev:2013usa}. The scale of the cosmological constant is determined by the graviton mass, and its small value is protected by the diffeomorphism symmetry in the massless limit \cite{deRham:2013qqa}. However, the cosmological solutions in the original dRGT theory are  unstable \cite{Gumrukcuoglu:2011zh, DeFelice:2012mx}.

The stability of cosmological solutions in massive gravity can be established in the framework of extensions of dRGT with additional degrees of freedom (see e.g. \cite{DeFelice:2013bxa,Hinterbichler:2016try} for reviews). On the other hand, perhaps the simplest extension that preserves the number of degrees of freedom of dRGT, without sacrificing the Lorentz invariance, is the Generalised Massive Gravity (GMG) theory \cite{deRham:2014lqa,deRham:2014gla}. In this setup, the translation invariance in the field space is broken by promoting the theory parameters to functions of the field-space invariant $\eta_{ab}\phi^a\phi^b$. This theory is part of a general class of massive gravity theories \cite{Gumrukcuoglu:2020utx} and provides stable cosmologies where all $5$ graviton polarisations propagate \cite{Kenna-Allison:2019tbu}.
In the context of a minimal proof-of-principle model where only one of the free parameters of dRGT varies, 
Ref.~\cite{Kenna-Allison:2020egn} showed that the generalised mass terms invoke a dynamical dark energy with an equation of state $w <-1$, sourcing a late-time accelerated expansion, while the scalar and tensor perturbations have potentially observable deviations from GR. The presence of an extra scalar mode contributing to the gravitational dynamics calls for a screening on short scales in order to satisfy local tests of gravity \cite{Will:2014kxa} via the Vainshtein mechanism.

The aim of the present paper is to study the non-linear perturbations in GMG around a cosmological background and investigate the nature of the Vainshtein screening mechanism. In Ref.~\cite{Kimura:2011dc} the Vainshtein mechanism was studied in the context of a cosmological background in Horndeski theory, which is the most general scalar-tensor theory that leads to second order equations of motion \cite{Horndeski:1974wa}. We follow a similar procedure here. We employ both perturbative and non-perturbative approaches and show that although there are similarities to scalar-tensor cosmologies, the square-root structure of the massive gravity potential leads to a new type of non-linear behaviour for the scalar mode. We also confirm the presence of screening which suppresses the modification of gravity below the Vainshtein radius.

The paper is organised as follows. In Sec.~\ref{sec2}, we briefly review GMG and introduce a setup for cosmological perturbations. In Sec.~\ref{sec3}, we derive covariant non-linear equations up to quadratic order around a cosmological background and show that the derivation procedure in \cite{Kimura:2011dc} cannot be directly applied due to the presence of an infinite series expansion. In Sec.~\ref{sec4}, we derive equations for spherically symmetric cosmological perturbations in order to take all relevant non-linear terms into account. In Sec.~\ref{sec:vainshtein-case1}, we numerically solve the non-linear equations and find asymptotic solutions at large and short distances. We conclude with Sec.~\ref{sec6} where we discuss our results.  The paper is complemented by Appendix~\ref{app:cosmoeq}, where we present the Einstein and St\"uckelberg equations up to quadratic order; and Appendix~\ref{app:mastereq}, where we show the explicit form of the St\"uckelberg and master equations for spherically symmetric perturbations.

\section{Generalised Massive Gravity}
\label{sec2}
In this section we present a brief review of the GMG action, derive the equations of motion and discuss the field configuration relevant for the subsequent sections.
\subsection{Set up}

The gravitational action consists of the Einstein-Hilbert term and the generalised mass terms \cite{deRham:2014gla}
\begin{equation}\label{model1}
S=\frac{M_p^2}{2}\int d^4 x \sqrt{-g}\left[R+2m^2\sum_{n=0}^4\alpha_n(\phi^a\phi_a)\;\mathcal{U}_n\left[\mathcal{K}\right]\right]+ \int d^4 x \sqrt{-g}\mathcal{L}_{m}\,,
\end{equation}
where $\mathcal{L}_m$ is the matter Lagrangian coupled minimally to the physical metric $g$. The graviton mass terms are given by the dRGT potentials $\mathcal{U}_n$ which are the elementary symmetric polynomials
\begin{align}\label{Ufunc}
    \mathcal{U}_0(\mathcal{K})&=1 \,\nonumber\\
    \mathcal{U}_1(\mathcal{K})&=[\mathcal{K}]\,\nonumber\\
    \mathcal{U}_2(\mathcal{K})&=\frac{1}{2!}\left([\mathcal{K}]^2-[\mathcal{K}^2]\right)\,\nonumber\\
    \mathcal{U}_3(\mathcal{K})&=\frac{1}{3!}\left([\mathcal{K}]^3-2[\mathcal{K}][\mathcal{K}]^2+2[\mathcal{K}^3]\right)\,\nonumber\\
    \mathcal{U}_4(\mathcal{K})&=\frac{1}{4!}\left([\mathcal{K}]^4-6[\mathcal{K}^2][\mathcal{K}]^2+8[\mathcal{K}^3][\mathcal{K}]-6[\mathcal{K}]^4 \right),
\end{align}
which are tuned such that the Boulware-Deser ghost is not excited \cite{deRham:2010ik,deRham:2010kj}. In (\ref{Ufunc}) square-brackets denote matrix trace operation and $\mathcal{K}$ is defined as
\begin{equation}
    \mathcal{K}^{\mu}_{\;\;\nu}\equiv\delta^{\mu}_{\;\;\nu}-\gamma^{\mu}_{\;\;\nu},
\end{equation}
where $\gamma$ is the matrix square-root of the $g^{-1}f$ tensor 
defined through
\begin{equation}
   \gamma^{\mu}_{\;\;\gamma} \gamma^{\gamma}_{\;\;\nu}=g^{\mu \gamma}f_{\gamma\nu }.
\end{equation}
The mass of the graviton is thus constructed as an interaction term between the physical metric $g$ and the fiducial metric $f$ defined in Eq.~\eqref{eq:f-defined}.
In GMG theory, the translation invariance in the St\"uckelberg field space is broken. As a result, one can promote the $\alpha_n$ parameters appearing in the dRGT potential term to be functions of the combination $\eta_{ab}\phi^a\phi^b$ without generating the Boulware-Deser ghost \cite{deRham:2014gla}.

To obtain the generalisation of the Einstein's equations, we vary the action (\ref{model1}) with respect to $g_{\mu \nu}$. The resulting equations of motion 
are
\begin{equation}
	\mathcal{E}^\mu_{\;\;\nu} \equiv G^{\mu}_{\;\;\nu} - \frac{1}{M_p^2}\,T^\mu_{\;\;\nu} - m^2\,\mathcal{Q}^\mu_{\;\;\nu}\,,
	\label{eq:einseqs}
\end{equation}
where $\mathcal{Q}^\mu_{\;\;\nu}$ is the effective stress-energy tensor arising from the mass term defined as
\begin{equation}
	\mathcal{Q}^{\mu}_{\;\;\nu} \equiv \sum_{n=0}^4\,\alpha_n(\phi^a\phi_a)\,\left(\delta^\mu_{\;\;\nu} \mathcal{U}_n - 2\,g^{\mu\rho} \,\frac{\delta \mathcal{U}_n}{\delta g^{\rho\nu}}\right)\,,
\end{equation}
and the energy-momentum tensor $T^{\mu}_{\;\;\nu}$ is given by
\begin{equation}
    T^{\mu}_{\;\;\nu}\equiv -\frac{2}{\sqrt{-g}}\frac{\delta}{\delta g_{\mu}^{\;\;\nu} }(\sqrt{-g}\mathcal{L}_m)\,.
\end{equation}
The variation of the dRGT potentials $\mathcal{U}_n$ can be found in Ref.\cite{Kenna-Allison:2020egn}. For the remainder of the paper, we consider a dust fluid as the matter content
\begin{equation}\label{enmomBG}
	T^\mu_{\;\;\nu} \equiv \rho \,u^\mu u_\nu\,,
\end{equation}
where $u^{\mu}$ is the 4-velocity and we assume 
that the energy-momentum tensor is covariantly conserved
\begin{equation}
	\nabla_\mu  T^\mu_{\;\;\nu} =0\,.
	\label{eq:enmomcons}
\end{equation}
Taking the divergence of (\ref{eq:einseqs}) assuming (\ref{eq:enmomcons}) and using the Bianchi identity yields,
\begin{equation}\label{stuckeom}
    \nabla_{\mu}\mathcal{Q}^{\mu}_{\;\;\nu}=0,
\end{equation}
which represents the equation of motion for the St\"uckelberg fields.

\subsection{Decomposition of the two metrics}
\label{sec:fiducial}
In this section we outline the metric decomposition that will be used in the discussion of the dynamics away from the cosmological background. Note at this time we make no assumption about the order of the perturbations.
\subsubsection{$g$-metric}
For the $g$ metric we choose to use the Newtonian gauge in an open FLRW background defined as,
\begin{equation}\label{gmet}
	ds^2 = -(1+2\,\Phi)dt^2 + a^2 (1-2\,\Psi)\Omega_{ij}dx^i dx^j\,,
\end{equation}
where $\Omega_{ij}$ is the 3-metric of a space-like hypersurface with constant negative curvature 
\begin{equation}
\Omega_{ij}dx^i dx^j=dx^2+dy^2+dz^2-\frac{\kappa (xdx+ydy+zdz)^2}{1+\kappa(x^2+y^2+z^2)}\,,
\end{equation}
where $\kappa$ is the absolute value of the negative spatial curvature, i.e. $\kappa=-|K|$. The choice of the open universe solution is imposed by the appearance of $\phi^a\phi_a$ in the GMG potential which only allows homogeneous and isotropic solutions with negative curvature \cite{Gumrukcuoglu:2020utx}. 

In the coming sections we will adopt a zero curvature limit for simplicity, under which $\Omega_{ij}$ reduces to the Euclidean three-space $\delta_{ij}$. In this limit, the metric (\ref{gmet}) becomes,
\begin{equation}
ds^2 = -(1+2\,\Phi)dt^2 + a^2 (1-2\,\Psi)\delta_{ij}dx^i dx^j.
\end{equation}
\subsubsection{$f$-metric}
For the background configuration in the St\"uckelberg sector, we use a variant of the non-Lorentz invariant gauge in Ref.~\cite{deRham:2014lqa}, given by \cite{Gumrukcuoglu:2011zh,Gumrukcuoglu:2020utx}
\begin{equation}
	\langle \phi^0 \rangle =f(t)\,\sqrt{1 +\kappa\,x^i\delta_{ij}x^j}\,,\qquad
	\langle \phi^i \rangle = \sqrt{\kappa}\,f(t)\,x^i\,.
\end{equation}
 In order to determine a convenient decomposition of perturbations, we introduce coordinate perturbations 
\begin{equation}
	x^\mu \to x^\mu + \Pi^a\,\delta^\mu_a\,.
	\label{eq:coordinate-transformation}
\end{equation}
At this point, we adopt the zero curvature scaling limit introduced in Ref.~\cite{deRham:2014lqa} to make the handling of the non-linear perturbation analysis easier. In particular we set
\begin{equation}
	f(t)=\frac{\alpha}{\sqrt{\kappa}}+\chi(t)\,,
\end{equation}
where $\alpha$ is a constant that is sensitive to the normalisation of the scale factor.\footnote{The value of $\alpha$ is not physical, but it is helpful for non-trivial checks of expressions, since it allows us to distinguish between physical $\alpha/a$ and the scale factor $a$ whose magnitude is unphysical.
}

Expanding for small curvature but not for small perturbations, the St\"uckelberg fields become 
\begin{align}
	\phi^0 &= \frac{\alpha}{\sqrt{\kappa}}+\chi(t+\Pi^0) + \mathcal{O}(\sqrt{\kappa})\,,\nonumber\\
	\phi^i &= 
	\alpha\, \left(x^i + \partial^i\,\Pi\right)+\mathcal{O}(\sqrt{\kappa})\,.
\end{align}
In GMG, we also need the following quantity: 
\begin{equation}
	\phi^a\eta_{ab}\phi^b = -\frac{\alpha^2}{\kappa} -\frac{2\,\alpha}{\sqrt{\kappa}}\,\chi(t+\Pi^0)+ \mathcal{O}(\kappa^0)\,,
\end{equation}
which diverges in the zero curvature limit. However, this quantity appears only in the arguments of unknown functions, therefore we can instead use the related quantity
\begin{equation}
	\tilde{\phi}^a\tilde{\phi}_a \equiv -\frac{\sqrt{\kappa}}{2\,\alpha}\,\left(\phi^a\phi_a+\frac{\alpha^2}{\kappa}\right) = \chi(t+\Pi^0)+ \mathcal{O}(\sqrt{\kappa})\,.
	\label{eq:PhiaPhia}
\end{equation}
Finally, the components of the fiducial metric \eqref{eq:f-defined}
in the limit $\kappa\to0$ are given by
\begin{align}\label{fmet}
	f_{00} &= 
	-\left[  \chi'(t+\Pi^0) (1+\dot\Pi^0)\right]^2+\alpha^2\,\partial^i\dot\Pi\,\partial_i \dot\Pi
	\,,\nonumber\\
	f_{0i} &= 
	-\left[  \chi'(t+\Pi^0)\right]^2 (1+\dot\Pi^0) \,\partial_i\Pi^0+\alpha^2\,\left[\partial_i\dot\Pi+\partial^k\dot\Pi\,\partial_k\partial_i\Pi\right]
	\,,\nonumber\\
	f_{ij} &= -\left[  \chi'(t+\Pi^0)\right]^2 \partial_i\Pi^0\,\partial_j\Pi^0 + \alpha^2\left(\delta_{ij}+2\,\partial_i\partial_j\Pi +\partial_i\partial^k\Pi\,\partial_k\partial_j\Pi\right)
	\,.
\end{align}
Note that we did not yet perform a perturbative expansion. In the next section, we implement these expressions and expand the equations of motion to the desired order in perturbations.

\section{Cosmological perturbations}
\label{sec3} 

In this section we perform cosmological perturbation analysis. We first discuss the background briefly and then derive the master equations for perturbations using the quasi-static approximation, keeping relevant non-linear terms for the study of the Vainshtein mechanism up to the second order. 

\subsection{Background}
To obtain the background equations of motion, we substitute the metric tensors (\ref{gmet}), (\ref{fmet}) and the energy-momentum tensor (\ref{enmomBG}) into the equations of motion (\ref{eq:einseqs}), 
while switching off all the perturbations.
The background Einstein equations take the following form,
\begin{align}\label{BGeins}
    &3H^2=\frac{\rho}{M_p^2} + m^2L \,,\nonumber\\
    &2\dot{H} =-\frac{\rho}{M_p^2}+m^2 (\ct-1)\CC_1\,,
\end{align}
where 
\begin{equation}
 \xi\equiv \frac{\alpha}{a}\,,\qquad
 \tilde{c} \equiv \frac{\dot\chi}{\xi}\,,\qquad
 H\equiv \frac{\dot{a}}{a}\,.
 \label{eq:defxr}
\end{equation}

In the above and the rest of the paper, we use the following combinations to replace $\alpha_1$, $\alpha_2$, $\alpha_3$ and $\alpha_4$:
\begin{align}
L(\chi,\xi) \equiv& -\alpha_0(\chi)+(3\,\xi - 4)\alpha_1(\chi) -3\,(\xi-2)(\xi-1)\alpha_2(\chi)+(\xi-4)(\xi-1)^2\alpha_3(\chi)+(\xi-1)^3\alpha_4(\chi)\,,\nonumber\\
\CC_1(\chi,\xi) \equiv & \,\xi\,\left[\alpha_1(\chi) +(3-2\xi)\,\alpha_2(\chi)+(\xi-3)(\xi-1)\,\alpha_3(\chi)+(\xi-1)^2\alpha_4(\chi)\right]\,,\nonumber\\
\CC_2(\chi,\xi) \equiv & \,2\,\xi^2 \left[
-\alpha_2(\chi)+(\xi-2)\,\alpha_3(\chi)+(\xi-1)\,\alpha_4(\chi)
\right]\,,\nonumber\\
\CC_3 (\chi,\xi)\equiv &\xi^3\,\left[\alpha_3(\chi)+\alpha_4(\chi)\right]\,.
\end{align}

In addition to the Einstein's equations, the matter conservation equation for the dust fluid \eqref{eq:enmomcons} implies
\begin{equation}
    \dot{\rho}+3H\rho=0.
\end{equation}
The equation of motion of the St\"uckelberg field can be derived using the above via the contracted Bianchi identity (\ref{stuckeom}) as
\begin{equation}
    \dot{L}-3\,H (\ct-1)\CC_1=0.
\end{equation}
In the perturbative study, we will use these background equations to remove $\dot{H}$, $\dot{\rho}$ and $L$.

In standard dRGT where there is no $\chi$ dependence, $\dot{L}\propto H\,\CC_1$, so the St\"uckelberg equation of motion imposes $\CC_1=0$. This is the source of the infinite strong coupling problem in dRGT, as the kinetic terms of vector and scalar perturbations are proportional to $\CC_1$ \cite{Gumrukcuoglu:2011zh}. In contrast, GMG avoids this strong coupling thanks to the varying coupling constants, which force $\CC_1$ away from zero \cite{Kenna-Allison:2019tbu,Kenna-Allison:2020egn}.

\subsection{Perturbations}

To begin the study of perturbations we assume all perturbation quantities to be of the same order in the gradient expansion $\Phi,\,\Psi,\,\Pi,\,\Pi^0 = \mathcal{O}(\epsilon) \ll 1$. We then count the number of spatial derivatives and keep the leading order terms in the gradient expansion. Note that the density perturbations $\delta\rho(t,{\bf x}) \equiv \rho(t,{\bf x})-\rho(t)= \mathcal{O}(\epsilon\,\partial_i^2)$ are enhanced compared with other variables. 
In order to compute the square root of the matrix $g^{-1} f$, we need to perform an expansion in terms of $\mathcal{O}(\epsilon\,\partial_i^2)$. We keep up to quadratic terms involving the second spatial derivatives of $\Phi,\,\Psi$ and $\Pi$. We present all equations in Appendix A. Here we only show three equations that are necessary to find the solutions for $\Phi,\,\Psi$ and $\Pi$. 

The relevant Einstein equations are given by
\begin{align}
	\label{eq:E00S}
	\delta\mathcal{E}^0_0 =& -\frac{\delta\rho}{M_p^2}-m^2\,\CC_1\,\partial^2\Pi + \frac{2}{a^2}\partial^2\Psi
	- \frac{m^2\CC_2}{4} \left[(\partial^2\Pi)^2-(\partial_i\partial_j\Pi)^2
	\right]\,, \\
	\label{eq:Etls}
	\mathcal{E}^{{\rm trless}} = & -\frac{2}{3\,a^2}\,\partial^2 \left( \Phi-\Psi+\frac{m^2a^2\,[2\,\CC_1+\CC_2(\ct-1)]}{2}\,\Pi\right) -\frac{m^2[\CC_2+2\,\CC_3(\ct-1)]}{12}\,[(\partial^2\Pi)^2-(\partial_i\partial_j\Pi)^2]\,,
\end{align}
where the first one is the time-time component and the second one is the traceless part of the space-space component.

We solve the temporal component of the St\"uckelberg equation \eqref{eq:stuckelbergT-perturbative}
for $\Pi^0$ and substitute it into the spatial component \eqref{eq:stuckilin}--\eqref{eq:stuckiquad}
to obtain the reduced equation:
\begin{align}
	\label{eq:nonlinearPi}
	\delta \mathcal{E}^{\rm St}_i =& \partial_i\left(
	\CC_1\,\Phi-[2\,\CC_1+\CC_2(\ct-1)]\,\Psi +\A_1\,a^2H^2\,\Pi
	\right)\,,
	\nonumber\\
	 &+
	\partial_j\Psi\,
	\left[\left(\frac{\CC_2}{2}+\CC_3(\ct-1)\right)\left(\partial_i\partial^j\Pi-\delta_i^j\partial^2\Pi\right)-[2\,\CC_1+\CC_2(\ct-1)]\,\partial_i\partial^j\Pi\right]
	\nonumber\\
	&-\partial_j\Phi\,
	\left[\frac{\CC_2}{2}\left(\partial_i\partial^j\Pi-\delta_i^j\partial^2\Pi\right)-\CC_1\,\partial_i\partial^j\Pi\right]+a^2\,H^2\,\partial_j\Pi\,
	\left[
	\A_2 \left(\partial_i\partial^j\Pi-\delta_i^j\partial^2\Pi\right)+\A_1 \partial_i\partial^j\Pi
	\right]
    \nonumber\\
	&+\frac{a^2H}{4\,\tilde{c}}
	\left(-2\,\CC_1+(\CC_2-2\,\CC_3)(\ct-1)+\frac{\CC_2^2(\ct^2-1)}{\CC_1} 
	+\frac{\CC_1(-2\,\dot{\CC}_1+\dot{\CC}_2)-\CC_2(\ct+1)\dot{\CC}_1}{H\,\CC_1}
\right)\partial_j\left(\partial_i\Pi\,\partial^j\dot\Pi-\partial_i\dot\Pi\,\partial^j\Pi\right)
	\,,
\end{align}
where $\A_1$ and $\A_2$ are defined in Eq.~\eqref{eq:defA1A2}.

In the Einstein equations, the non-linear terms of the second derivative of $\Pi$ takes the same form as those found the Horndeski theory \cite{Kimura:2011dc}. On the other hand, the additional equation for $\Pi$ obtained from the St\"uckelberg equation (after taking the divergence of (\ref{eq:nonlinearPi})) contains the time derivative as well as the third spatial derivative of $\Pi$. This is reminiscent of the DHOST theories \cite{Crisostomi:2017lbg, Crisostomi:2017pjs}. However, in GMG, the non-linear terms in this equation are not truncated at finite order. This is because the solution for $\Pi^0$ contains the non-linear terms of the second derivative of $\Pi$ at infinite order if expanded. In order to demonstrate this point and obtain non-perturbative solutions, we will consider static and spherically symmetric perturbations in the following section. 

\section{Spherically symmetric perturbations}
\label{sec4} 

Although the perturbative study in the previous section is useful to unveil the structure of non-linear interactions, it is limited by our inability to compute the square root matrix without performing a perturbative expansion. To obtain interactions at arbitrary order, we consider spherically symmetric deviations from FLRW in a non-perturbative setup.

\subsection{Set-up}
For the St\"uckelberg field configuration, we proceed in the same way as prescribed in Sec.\ref{sec:fiducial}, also utilising the zero-curvature scaling limit:
\begin{align}
\phi^0 =& \frac{\alpha}{\sqrt{\kappa}}+\chi(t)+\delta\chi(t,r)\,,\nonumber\\
\phi^i =& \alpha\,\big(x^i +\partial^i\Pi(t,r)\big)\,,
\end{align}
where the indices in the field space correspond to Cartesian coordinates and we defined $\delta\chi(t,r) \equiv \chi(t+\Pi^0) - \chi(t)$. Note that $r$ is the radial coordinate $r^2\equiv \delta_{ij} x^i x^j$. 

In this section,
we do not assume that the perturbations are small. Analogous to Eq.\eqref{eq:PhiaPhia}, we find that the norm of the St\"uckelberg vector is
\begin{equation}
 \tilde{\phi}^a\tilde{\phi}_a = \chi(t)+\delta\chi(t,r)\,.
\end{equation}
We can now compute the fiducial metric. 
We start by first determining it in
Cartesian coordinates $(t,x,y,z)$
\begin{align}
f_{00}=&-(\dot{\chi}+\delta\dot\chi)^2+\alpha^2 (\dot\Pi')^2\,,\nonumber\\
f_{0i} = & \frac{x^i}{r}\left[-(\dot{\chi}+\delta\dot\chi)\,\delta\chi'+\alpha^2\,\dot\Pi'\left(1+\Pi''\right)\right]\,,\nonumber\\
f_{ij} = & -\frac{x^ix^j}{r^2}\,(\delta\chi')^2 + \alpha^2 \left[ \delta_{ij} +\frac{2}{r}\,\left(\delta_{ij}-\frac{x^ix^j}{r^2}\right)\,\Pi' + \frac{2\,x^ix^j}{r^2}\,\Pi'' + \frac{1}{r^2}\left(\delta_{ij}-\frac{x^ix^j}{r^2}\right)\,(\Pi')^2+\frac{x^ix^j}{r^2}\,(\Pi'')^2
\right]\,.
\end{align}
Then, to analyse the spherically symmetric perturbations we 
transform to spherical coordinates $(t,r,\theta,\varphi)$ using the invariance of the line element, which yields
\begin{align}
f_{\mu\nu}dx^\mu dx^\nu =& 
\bar{f}_{\mu\nu}d\bar{x}^\mu d\bar{x}^\nu \nonumber\\
=& \bar{f}_{tt} \,dt^2 + \bar{f}_{rr}\,dr^2 + 2\,\bar{f}_{tr}dt\,dr+ \bar{f}_{\theta\theta} d\Omega_3^2\nonumber\\
=&[-(\dot{\chi}+\delta\dot\chi)^2+\alpha^2(\dot\Pi')^2]\,dt^2 +[-(\delta\chi')^2+\alpha^2(1+\Pi'')^2]\,dr^2
\nonumber\\
&
+2\,[-(\dot\chi+\delta\dot\chi)\delta\chi'+\alpha^2
\dot{\Pi'}
(1+\Pi'')]\,dt\,dr+\alpha^2(r+\Pi')^2d\Omega_{2}^2,
\end{align}
where $d\Omega_{2}^2$ is the 
metric of a 2--sphere.
For the physical metric we use the longitudinal gauge and again avoid using any perturbative expansion
\begin{equation}
g_{\mu\nu}dx^\mu dx^\nu = -(1+2\,\Phi)\,dt^2 + a^2(1-2\,\Psi)(dr^2+r^2d\Omega_{2}^2)\,.
\end{equation}
We are now in a position to compute the square-root matrix $\gamma \equiv \sqrt{g^{-1}f}$ which appears in the graviton mass term. We begin by calculating 
$\gamma^2$
as follows
\begin{equation}
\gamma^2 = g^{-1}f=\left(
\begin{array}{llll}
\frac{\bar{f}_{tt}}{g_{tt}} & 
\frac{\bar{f}_{tr}}{g_{tt}} & 0 & 0 \\
\frac{\bar{f}_{tr}}{g_{rr}} & \frac{\bar{f}_{rr}}{g_{rr}} & 0 &0\\
0 & 0 & \frac{\bar{f}_{\theta\theta}}{g_{\theta \theta}} & 0\\
0 & 0 & 0 & \frac{\bar{f}_{\varphi\varphi}}{g_{\varphi \varphi}} 
\end{array}
\right)\,.
\end{equation}
Since the off-diagonal block is $2\times2$, the square-root tensor can be computed using the following identity for 2 dimensional matrices \cite{Gratia:2012wt}
\begin{equation}\label{ident}
\sqrt{\mathcal{X}} = \frac{1}{[{\rm Tr}(\mathcal{X})+2\,\sqrt{{\rm Det}(\mathcal{X})}]^{1/2}}\,\left(\mathcal{X} + \sqrt{{\rm Det}(\mathcal{X})}\,I_2\right)\,,
\end{equation}
where $I_2$ is the $2\times 2$ identity matrix and $\mathcal{X}$ is a generic $2\times 2$ matrix. Using (\ref{ident}) we can write the components of $\gamma$ as follows,
\begin{align}
\gamma^{\,t}_{\;\;\;t} =&  \frac{f_{tt}+ g_{tt}\sqrt{\frac{f_{tt}\,f_{rr}-f_{tr}^2}{g_{tt}g_{rr}}}}{g_{tt} \left[
\frac{f_{tt}}{g_{tt}} + \frac{f_{rr}}{g_{rr}} +2\,\sqrt{\frac{f_{tt}f_{rr}-f_{tr}^2}{g_{tt}g_{rr}}}
\right]^{1/2}}\,,\nonumber\\
\gamma^{\,t}_{\;\;\;r} =&  \frac{
f_{tr}
}{g_{tt} \left[
\frac{f_{tt}}{g_{tt}} + \frac{f_{rr}}{g_{rr}} +2\,\sqrt{\frac{f_{tt}f_{rr}-f_{tr}^2}{g_{tt}g_{rr}}}
\right]^{1/2}}\,,\nonumber\\
\gamma^{\,r}_{\;\;\;t} =&  \frac{
f_{tr}
}{g_{rr} \left[
\frac{f_{tt}}{g_{tt}} + \frac{f_{rr}}{g_{rr}} +2\,\sqrt{\frac{f_{tt}f_{rr}-f_{tr}^2}{g_{tt}g_{rr}}}
\right]^{1/2}}\,,\nonumber\\
\gamma^{\,r}_{\;\;\;r} =&  \frac{f_{rr}+ g_{rr}\sqrt{\frac{f_{tt}\,f_{rr}-f_{tr}^2}{g_{tt}g_{rr}}}}{g_{rr} \left[
\frac{f_{tt}}{g_{tt}} + \frac{f_{rr}}{g_{rr}} +2\,\sqrt{\frac{f_{tt}f_{rr}-f_{tr}^2}{g_{tt}g_{rr}}}
\right]^{1/2}}\,,\nonumber\\
\gamma^{\,\theta}_{\;\;\;\theta} =&  \gamma^{\,\varphi}_{\;\;\;\varphi}=\frac{\sqrt{f_{\theta\theta}}}{r\,\sqrt{g_{rr}}}\,,
\end{align}
where we suppressed the overbars for clarity of presentation. 
As we mentioned above, instead of a perturbative treatment, in this section we wish to keep track of self-interactions of $\Pi$ at arbitrary order. To simplify this exercise, we assume that all perturbations (except density) are of the same order and small. The difference is that in the small scales that we consider, the spatial derivatives enhance them and in particular, second derivatives of perturbations are of order 1. This allows us to keep track of $\Pi$ interactions while allowing Einstein's equations to be consistent. Quantitatively, we take
\begin{equation}
\Phi\,, \Psi\,, \delta\chi\,,\Pi\,,v \sim \mathcal{O}(\epsilon^2)~~\,,~~ \delta\rho\sim\mathcal{O}(1) ~~~{\rm and}~~~
 r \sim \mathcal{O}(\epsilon)\,,
 \label{eq:nonlinear-expansion}
\end{equation}
such that $\Pi'\sim\mathcal{O}(\epsilon)$, $\Pi''\sim \mathcal{O}(\epsilon^0)$ and similarly for other perturbations.

When discussing the Vainshtein mechanism, it is better to use the physical coordinate ${\tilde r}=a r$ rather than the comoving coordinate $r$. In addition, the dependence of the scale factor in all equations derived in this section can be absorbed by defining ${\tilde \Pi} \equiv a^2 \Pi$ when we use the physical coordinate. 
	However, we use the comoving coordinate throughout this paper since we set $a=1$ and the comparison with the second-order perturbation equations \eqref{eq:E00S}-\eqref{eq:nonlinearPi} is clear.

\subsection{St\"uckelberg equation of motion}
We now compute the effective energy-momentum tensor of the mass term $\mathcal{Q}^{\mu}_{\;\;\nu}$, and from its conservation calculate the St\"uckelberg equation as 
outlined
in Appendix \ref{sec:stuckelbergeq}. The temporal component is formally 
\begin{equation}
 \mathcal{E}^{\rm St}_0 \equiv \nabla_\mu \mathcal{Q}^{\mu}_{\;\;t}\,.
\end{equation}
At order $\epsilon^0$ this equation can be integrated once, similar to the perturbative case. By defining
\begin{equation}
 \frac{1}{r^2}\partial_r \delta\mathcal{E}^{{\rm St}\,r}_0 = \mathcal{E}^{\rm St}_0\,,
\end{equation}
we can integrate the temporal St\"uckelberg equation to obtain
\begin{align}
\delta\mathcal{E}^{{\rm St}\,r}_0 =& r^3 \Bigg\{
\left[\mathcal{C}_1+\mathcal{C}_2\,\frac{\Pi'}{r}+\mathcal{C}_3 \left(\frac{\Pi'}{r}\right)^2\right]
\left[-\frac{\tilde{c}}{\tilde{c}+1+\Pi''}\,\left(\frac{\delta\chi'}{a^2\,\xi\,r}+\frac{\dot\Pi'}{r}\right)+H\,\left(\tilde{c}-1-\frac{\Pi'}{r}\right)\right] 
\nonumber\\
&\qquad- \left[\dot{\mathcal{C}}_1\frac{\Pi'}{r} + \frac{\dot{\mathcal{C}}_2}{2}\left(\frac{\Pi'}{r}\right)^2+\frac{\dot{\mathcal{C}}_3}{3}\,\left(\frac{\Pi'}{r}\right)^3+\frac{\dot{L}}{3}\right]
\Bigg\}\,.
\label{eq:ST0integrated}
\end{align}
The integrated St\"uckelberg equation can be trivially solved for $\delta\chi'$:
\begin{align}
 \frac{\delta\chi'}{a^2\,\xi\,r} = -\frac{\dot\Pi'}{r}+\frac{\tilde{c}+1+\Pi''}{\tilde{c}}
 \left[
 H\,\left(\tilde{c}-1-\frac{\Pi'}{r}\right)
 -\frac{\dot{\mathcal{C}}_1\frac{\Pi'}{r} + \frac{\dot{\mathcal{C}}_2}{2}\left(\frac{\Pi'}{r}\right)^2+\frac{\dot{\mathcal{C}}_3}{3}\,\left(\frac{\Pi'}{r}\right)^3+\frac{\dot{L}}{3}}
{\mathcal{C}_1+\mathcal{C}_2\,\frac{\Pi'}{r}+\mathcal{C}_3 \left(\frac{\Pi'}{r}\right)^2}
 \right].
 \label{eq:dchisol}
\end{align}

We can now compare this result with the perturbative solution for $\Pi^0$ in Eq.~\eqref{eq:pi0sol}. Since perturbatively, we have $\delta\chi = \xi\,\tilde{c}\,\Pi^0$, the comparison is straightforward. By using the background St\"uckelberg equation $\dot{L}=3\,(\tilde{c}-1)\,H\,\mathcal{C}_1$ and the Friedmann equation, we can match all the terms in Eq.~\eqref{eq:pi0sol}. When we take into account the contribution from the spherical coordinates, i.e. $\partial_i\partial^i \Pi = \Pi'' + \frac{2}{r}\,\Pi'$, we find that the non-perturbative computation is in agreement with the perturbative one. 

Finally, we look at the spatial component of the St\"uckelberg equation. Using the solution for $\delta\chi'$ in Eq.\eqref{eq:dchisol}, its time and spatial derivates, we reduce the radial St\"uckelberg equation to the following form
\begin{align}
 \mathcal{E}^{\rm St}_r \equiv \nabla_\mu \mathcal{Q}^{\mu}_{\;\;r}
=& r\,(1+\Pi'')\Bigg\{
 \frac{\Phi'}{r} \left[\mathcal{C}_1+\mathcal{C}_2 \frac{\Pi'}{r}+\mathcal{C}_3 \left(\frac{\Pi'}{r}\right)^2\right]
-\frac{\Psi'}{r} \left[[2\,\mathcal{C}_1+\mathcal{C}_2(\tilde{c}-1)]+
[\mathcal{C}_2+2\,\mathcal{C}_3(\tilde{c}-1)]\frac{\Pi'}{r}
\right] 
\nonumber\\
& \qquad\qquad\qquad + \frac{\sum_{i=0}^7\mathcal{D}_i \left(\frac{\Pi'}{r}\right)^i}{\left(
\mathcal{C}_1+\mathcal{C}_2\frac{\Pi'}{r}+\mathcal{C}_3\frac{(\Pi')^2}{r^2}
\right)^2}\Bigg\}\,,
\label{eq:mastereq-formal}
 \end{align}
where $\mathcal{D}_i$ are functions of time, which we do not present here. The explicit form of the simplified equation is presented in Eq.\eqref{eq:mastereq-explicit} in Appendix \ref{app:mastereq}.

\subsection{Einstein's equations}

The time-time component of the Einstein equations \eqref{eq:einseqs} is computed as
\begin{align}
 \mathcal{E}^0_{\;\;\;0} =& -3\,H^2+m^2L+\frac{\rho+\delta\rho}{M_p^2}-\frac{2}{a^2}\left(\Psi''+\frac{2\,\Psi'}{r}\right)
 \nonumber\\
 & 
 +m^2\left[
 \left(2\,\mathcal{C}_1+\frac{\mathcal{C}_2}{2}\,\frac{\Pi'}{r}\right) \frac{\Pi'}{r} + \left(\mathcal{C}_1+\mathcal{C}_2\,\frac{\Pi'}{r} +\mathcal{C}_3\,\,\frac{(\Pi')^2}{r^2}\right)\Pi''
 \right]\,.
\end{align}

For the spatial components, there are only two independent equations $\mathcal{E}^r_{\;\;\;r}$ and $\mathcal{E}^\theta_{\;\;\;\theta}=\mathcal{E}^\varphi_{\;\;\;\varphi}$. To be able to compare with the perturbative calculation later, we decompose these into the trace and traceless parts via
\begin{equation}
 \mathcal{E}^{\rm tr}\equiv \frac{1}{3}\,\mathcal{E}^{k}_{\;\;\;k}\,,\qquad
 \mathcal{E}^{\rm trless}\equiv \lambda^j_{\;\;\;i}\left(\mathcal{E}^{i}_{\;\;\;j}-\frac{\delta^i_j}{3}\,\mathcal{E}^{k}_{\;\;\;k}\right)\,,
\end{equation}
where $\lambda^i_{\;\;\;j}={\rm diag}(1/2,-1,-1)$ is the inverse of a traceless diagonal matrix. In terms of the components, these are
\begin{equation}
\mathcal{E}^{\rm tr} = \frac{1}{3}\left(\mathcal{E}^r_{\;\;\;r}+2\,\mathcal{E}^\theta_{\;\;\;\theta}\right)\,,\qquad
\mathcal{E}^{\rm trless} = \mathcal{E}^r_{\;\;\;r}-\mathcal{E}^\theta_{\;\;\;\theta}\,,
\end{equation}
or,
\begin{align}
\mathcal{E}^{\rm tr} = & -3\,H^2-2\,\dot{H}+m^2L + m^2(\tilde{c}-1)\,\mathcal{C}_1
+\frac{2}{3\,a^2}\,\left(-\Psi''-\frac{2\,\Psi'}{r}+\Phi''+\frac{2\,\Phi'}{r}\right)
\nonumber\\
&
+\frac{m^2[2\,\mathcal{C}_1+\mathcal{C}_2(\tilde{c}-1)]}{3}\left(\Pi''+2\,\frac{\Pi'}{r}\right)
+\frac{m^2[\mathcal{C}_2+2\,\mathcal{C}_3(\tilde{c}-1)]}{6}\,\frac{\Pi'}{r}\left(2\,\Pi''+\frac{\Pi'}{r}\right)
\label{eq:EINTR}
\\
 \mathcal{E}^{\rm trless} =& \frac{1}{a^2}\left(\Psi''-\frac{\Psi'}{r}-\Phi''+\frac{\Phi'}{r}\right)-\frac{m^2}{2}\left[2\,\mathcal{C}_1+\mathcal{C}_2(\tilde{c}-1)+[\mathcal{C}_2+2\,\mathcal{C}_3(\tilde{c}-1)]\frac{\Pi'}{r}\right]\left(\Pi''-\frac{\Pi'}{r}\right)\,.
 \label{eq:EINTRLESS}
\end{align}

\subsection{Newtonian potentials}

In this subsection we outline the method taken to obtain the solutions for the two metric potentials. We define the total mass perturbation as
\begin{equation}
\delta M \equiv \int d^3r \sqrt{\det(g_{ij})}\,\delta\rho
=4\,\pi\,a^3\int r^2 dr (1-2\,\Psi)^{3/2}\,\delta\rho \simeq
4\,\pi\,a^3\int r^2 dr \,\delta\rho \,,
\end{equation}
where in the last step we took the dominant term in the expansion \eqref{eq:nonlinear-expansion}.
We can now integrate the 00 component of the metric equations of motion:
\begin{equation}
a^3M_p^2\int d^3r \,\mathcal{E}^0_{\;\;\;0}
=\frac{r^3a^3}{6\,G}\,\left[
-3\,H^2+m^2L+\frac{\rho}{M_p^2}
+\frac{6\,G\,\delta M}{r^3a^3} -\frac{6}{a^2}\,\frac{\Psi'}{r}
+3\,m^2\left(
\mathcal{C}_1\frac{\Pi'}{r}+\frac{\mathcal{C}_2}{2}\frac{(\Pi')^2}{r^2}+\frac{\mathcal{C}_3}{3}\,\frac{(\Pi')^3}{r^3}\right)
\right]+\mathcal{I}_1\,,
\end{equation}
where $\mathcal{I}_1=\mathcal{I}_1(t)$ is an integration constant and $G=(8\,\pi\,M_p^2)^{-1}$ is Newton's constant. Since in the linear regime the 00 equation reduces to the Friedmann equation, the time dependent function $\mathcal{I}_1$ is forced to vanish.
Solving for $\Psi'$, we find
\begin{equation}
\frac{\Psi'}{r} = \frac{a^2}{6}\,\left(-3\,H^2+m^2L+\frac{\rho}{M_p^2}\right)+\frac{G\,\delta M}{r^3 a}+\frac{m^2a^2}{2}\,\left(
\mathcal{C}_1\frac{\Pi'}{r}+\frac{\mathcal{C}_2}{2}\frac{(\Pi')^2}{r^2}+\frac{\mathcal{C}_3}{3}\,\frac{(\Pi')^3}{r^3}\right)\,.
\label{eq:solpsi'}
\end{equation}

We next determine $\Phi'$ by integrating the traceless equation:
\begin{equation}
\int \mathcal{E}^{\rm trless} \,\frac{ dr}{r}
=
\frac{1}{a^2}\left(\frac{\Psi'}{r}-\frac{\Phi'}{r}\right)
-\frac{m^2}{4}\left(
2\,[2\,\mathcal{C}_1+\mathcal{C}_2(\tilde{c}-1)]+[\mathcal{C}_2+2\,\mathcal{C}_3(\tilde{c}-1)]\frac{\Pi'}{r}\right)\frac{\Pi'}{r}\,,
\end{equation}
where we fixed the integration constant to vanish 
by requiring consistency with 
the linear regime. Solving this equation and using \eqref{eq:solpsi'}, we obtain
\begin{equation}
\frac{\Phi'}{r} = \frac{a^2}{6}\,\left(-3\,H^2+m^2L+\frac{\rho}{M_p^2}\right)+\frac{G\,\delta M}{r^3 a}-\frac{m^2a^2}{2}\,\left(
[\mathcal{C}_1+\mathcal{C}_2(\tilde{c}-1)]\,\frac{\Pi'}{r}+\mathcal{C}_3 (\tilde{c}-1)\,\frac{(\Pi')^2}{r^2}-\frac{\mathcal{C}_3}{3}\,\frac{(\Pi')^3}{r^3}\right)\,.
\label{eq:solphi'}
\end{equation}

Using these solutions (and their radial derivative) back in the trace equation \eqref{eq:EINTR}, we find an expression that is purely time dependent:
\begin{equation}
\mathcal{E}^{\rm tr}\Big\vert_{\rm reduced} = -3\,H^2-2\,\dot{H}+m^2L + m^2(\tilde{c}-1)\,\mathcal{C}_1\,,
\label{eq:isolatedHdot}
\end{equation}
which simply is one of the background equations in the perturbative calculation. 

\section{Vainshtein Radius - Transition into the Non-linear regime}
\label{sec:vainshtein-case1}

The goal of this section is to study the solutions for $\Pi'$. Using the solutions for the Newtonian potentials \eqref{eq:solpsi'}, \eqref{eq:solphi'} in the spatial St\"uckelberg equation \eqref{eq:mastereq-explicit}, we derive the master equation for $\Pi'$. The explicit form of this equation is given in \eqref{eq:mastereq-reduced}. 

\subsection{Assumptions and approximations}

Neglecting the trivial solution $\Pi''=-1$, Eq.~\eqref{eq:mastereq-reduced} is a ninth order polynomial equation of $\Pi'$. 
In order to study the full solutions, we numerically solve the equation. For this, we use several approximations:
\begin{enumerate}
 \item In our short distance expansion, the background and perturbations are intertwined and cannot be isolated, except for the acceleration equation \eqref{eq:isolatedHdot}. Therefore, to evaluate the master equation in the present Universe, we assume that we can separate the background and all background dynamics is dictated by the background equations in the perturbative treatment.
 \item For concreteness, we fix the mass functions $\alpha_n(\phi^a\phi_a)$ according to the minimal model of Ref.\cite{Kenna-Allison:2020egn}, where only $\alpha_2$ is non-constant and varies linearly,
 \begin{align}
 	\alpha_0(\phi^a\phi_a) = \alpha_1(\phi^a\phi_a)=\alpha_3(\phi^a\phi_a)=0\,,\qquad
 	\alpha_2(\phi^a\phi_a)= 1+ 10^{-4}\,H_0^2\,\phi^a\phi_a\,,\qquad
 	\alpha_4(\phi^a\phi_a)= 0.8\,,
 \label{model}
 \end{align}
which corresponds to 
$Q=10^4 q =1$
in the notation of \cite{Kenna-Allison:2020egn}. Requiring an early dRGT evolution, we fix
\begin{equation}
	m = 1.63\,H_0\,,
\end{equation}
such that the Friedmann equation is satisfied. We thus proceed as in Ref.\cite{Kenna-Allison:2020egn} and determine the full evolution of all background quantities. In the present paper, we evaluate all time-dependent quantities today by setting $a=1$, and we use the value of the sun $M_{\odot}\sim 10^{33} {\rm g}$ for the total mass perturbation, that is
\begin{equation}
	\delta M = \frac{10^{-23}}{G\,H_0}\,.
\end{equation} 

 \item In the non-perturbative calculation, we used the zero-curvature limit of Ref.\cite{deRham:2014lqa}. However, this limit is incompatible with the full evolution of the background. In the minimal model, the linear dependence requires an evolution very close to the $dRGT$ background, with a constant $\xi$. However in the zero-curvature limit $\xi$ is nothing but $1/a$, thus the early evolution cannot describe a matter dominated universe. 

 To circumvent this inconsistency, we reintroduce a curvature with density parameter 
$\Omega_{\kappa,0} =3 \times 10^{-3}$
  and evolve the background equations in an open Universe background. We then evaluate the time dependent coefficients in the master equation at $t=t_0$, where the effect of curvature is neglibigle. In numerical computation, we also fix the present matter density parameter as $\Omega_{{\rm m}0}=0.3$.
\end{enumerate}

\subsection{Numerical solution of the master equation}
Before discussing the numerical results, we first estimate the linear solution where $\Pi'$ and $\delta M$ can be treated as perturbations. Expanding the master equation \eqref{eq:mastereq-reduced}, the solution is 
\begin{align}
\Pi'_L =& -\frac{G\,\delta M 
	{\tilde c}^2\left[\mathcal{C}_1+\mathcal{C}_2(\tilde{c}-1)\right]
}{a^3r^2}
\Bigg[
\frac{\tilde{c}\,[\mathcal{C}_1-\mathcal{C}_2\tilde{c}^2(\tilde{c}-1)]\rho}{2\,M_p^2}
+\frac{m^2\,\tilde{c}\,\mathcal{C}_1\,\left[
\mathcal{C}_1(2\,\tilde{c}+1)+\mathcal{C}_2(\tilde{c}-1)(3\,\tilde{c}-1)
\right]}{2}
\nonumber\\
&\qquad\qquad\qquad\qquad\qquad\qquad
+\dot{\tilde{c}}\,\dot{\mathcal{C}}_1-\tilde{c}\,\ddot{\mathcal{C}}_1
+\left(-2(2\,\tilde{c}-1)\,\dot{\mathcal{C}}_1+(\tilde{c}-1)\tilde{c}\,\dot{\mathcal{C}}_2+(\mathcal{C}_1+\mathcal{C}_2)\dot{\tilde{c}}\right)H
\nonumber\\
&\qquad\qquad\qquad\qquad\qquad\qquad
-\left[
\mathcal{C}_1(4\,\tilde{c}-1)-\mathcal{C}_2(\tilde{c}-1)(3\,\tilde{c}-2)\right]\,H^2
\nonumber\\
&\qquad\qquad\qquad\qquad\qquad\qquad
+(\tilde{c}+1)\,\mathcal{C}_1\left(-\frac{\mathcal{C}_2}{\mathcal{C}_1}\,(\tilde{c}-1)\,H+\frac{\dot{\mathcal{C}}_1}{\mathcal{C}_1}\right)^2
\Bigg]^{-1}\,.
\label{eq:pi'linear}
\end{align}
This estimate allows us to determine the relevant solution out of the nine roots of the master equation. Since the linear solution of $\Pi_L'$ is proportional to $r^{-2}$, the scalar force induces scale-independent modification of the effective gravitational constant in the linear regime 
as one can see after substituting \eqref{eq:pi'linear} into \eqref{eq:solpsi'} and \eqref{eq:solphi'}.

\begin{figure}[h!]
	\centering
	\includegraphics[width=\columnwidth]{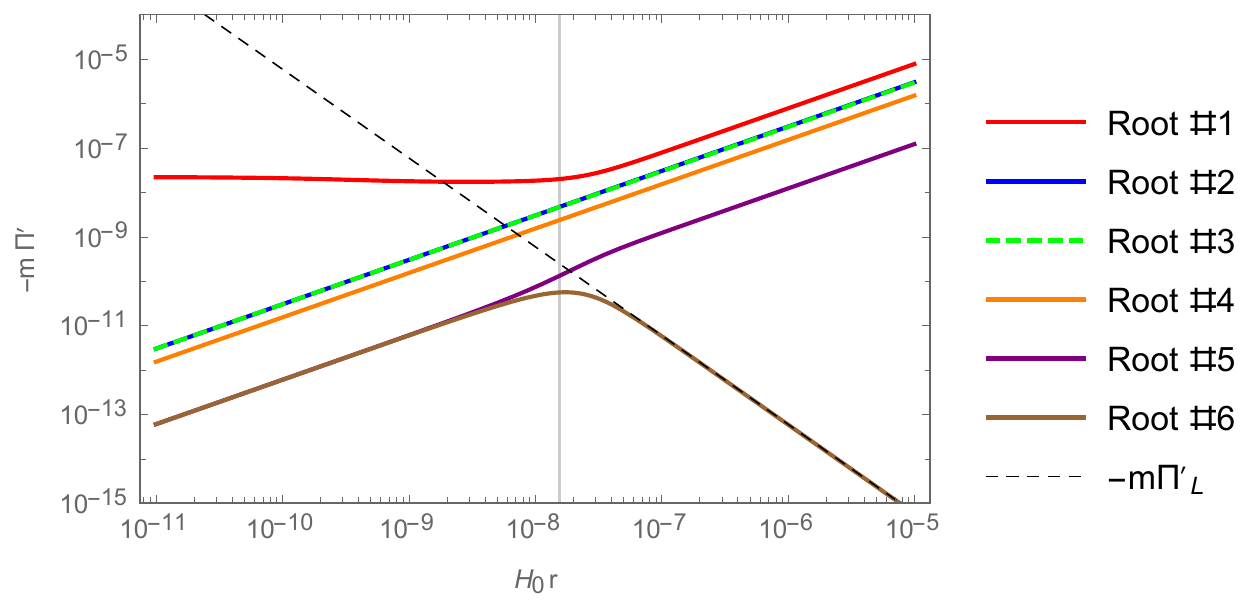}
	\caption{The plot of the six relevant roots of the master equation \eqref{eq:mastereq-reduced} as a function of radial distance. The dashed black line represents the analytic estimate for the linear solution \eqref{eq:pi'linear}.}
	\label{fig:allroots}
\end{figure}

Evaluating the master equation today, we determine the nine roots as a function of radial distance. We discard two complex roots and one that does not match the sign of the linear solution. The remaining six roots are shown in Fig.~\ref{fig:allroots}. We find that only one of the roots follows the analytic estimate \eqref{eq:pi'linear} in the linear regime.

\subsection{Analytic estimates for the non-linear regime}

We now calculate an analytic estimate for the non-linear regime. Our numerical results indicate three distinct behaviours of the solutions. 

\subsubsection{Constant solution}

Expanding the master equation \eqref{eq:mastereq-reduced} at small $r$ for constant $\Pi'$, we find the first non-linear solution as:
\begin{equation}
\Pi'_{NL1} = -\left(\frac{6\,G\,\delta M}{m^2 a^3\,\mathcal{C}_3}\right)^{1/3}\,.
\label{eq:pi'NL1}
\end{equation}

\begin{figure}[h!]
	\centering
	\includegraphics[width=0.6\columnwidth]{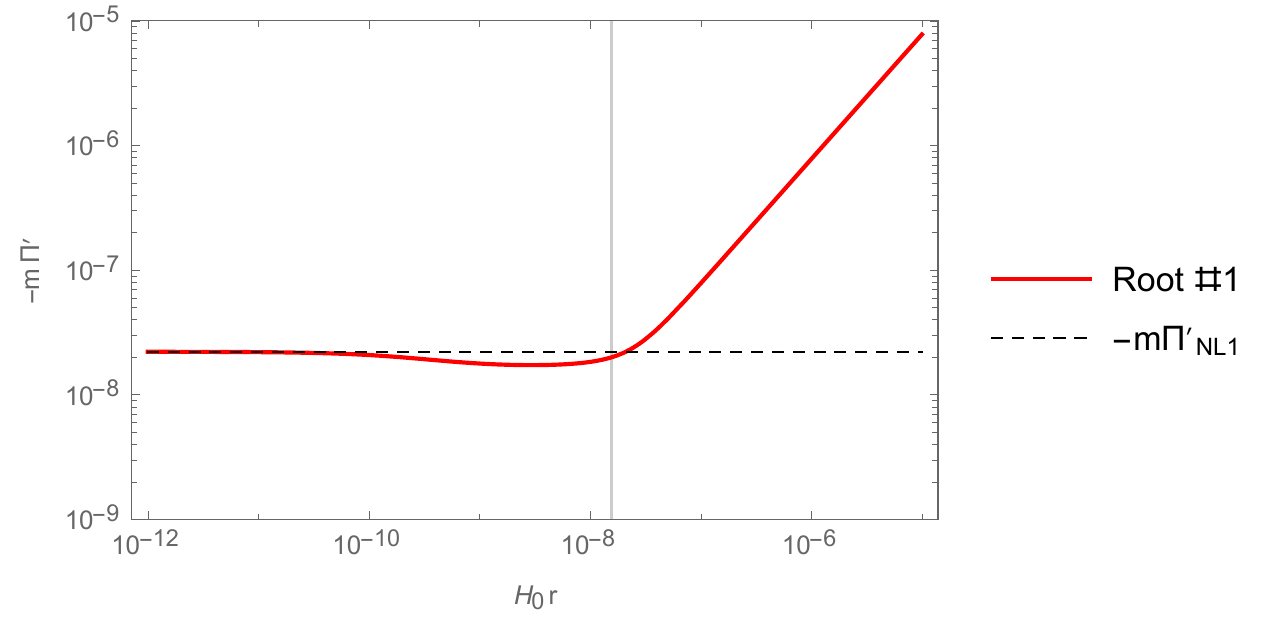}
	\caption{The root that matches the constant non-linear behaviour 
	\eqref{eq:pi'NL1}. The vertical line shows the radius where this solution coincides with the linear solution \eqref{eq:pi'linear}.}
 	\label{fig:PLOTNL1}
\end{figure}
In Figure~\ref{fig:PLOTNL1} we show that the root \#1 at short distances matches this non-linear solution, but fails to recover the linear behaviour at large distances. 

\subsubsection{Mass independent solution}

The second type of non-linear solution for $\Pi'$ depends linearly on $r$ in the non-linear regime. By considering  constant $\Pi'/r$, we can expand the master equation at short distances to the following quadratic equation:
\begin{equation}
 \mathcal{C}_1+\mathcal{C}_2(\tilde{c}-1)+2\,\mathcal{C}_3(\tilde{c}-1)\,\frac{\Pi_{NL2}'}{r} -\mathcal{C}_3\,\left(\frac{\Pi_{NL2}'}{r}\right)^2 = 0\,,
\end{equation}
which can be solved by
\begin{equation}
\Pi'_{NL2\pm}= r\,(\tilde{c}-1)\left(1 \pm \sqrt{1+\frac{\mathcal{C}_2}{\mathcal{C}_3(\tilde{c}-1)}+\frac{\mathcal{C}_1}{\mathcal{C}_3(\tilde{c}-1)^2}}\right)\,.
\label{eq:pi'NL2}
\end{equation}
We note that this solution is independent of the total mass perturbation $\delta M$, 
and is similar to the solution found in the context of Horndeski theory \cite{Kimura:2011dc}.

\begin{figure}[h!]
	\centering
	\includegraphics[width=0.6\columnwidth]{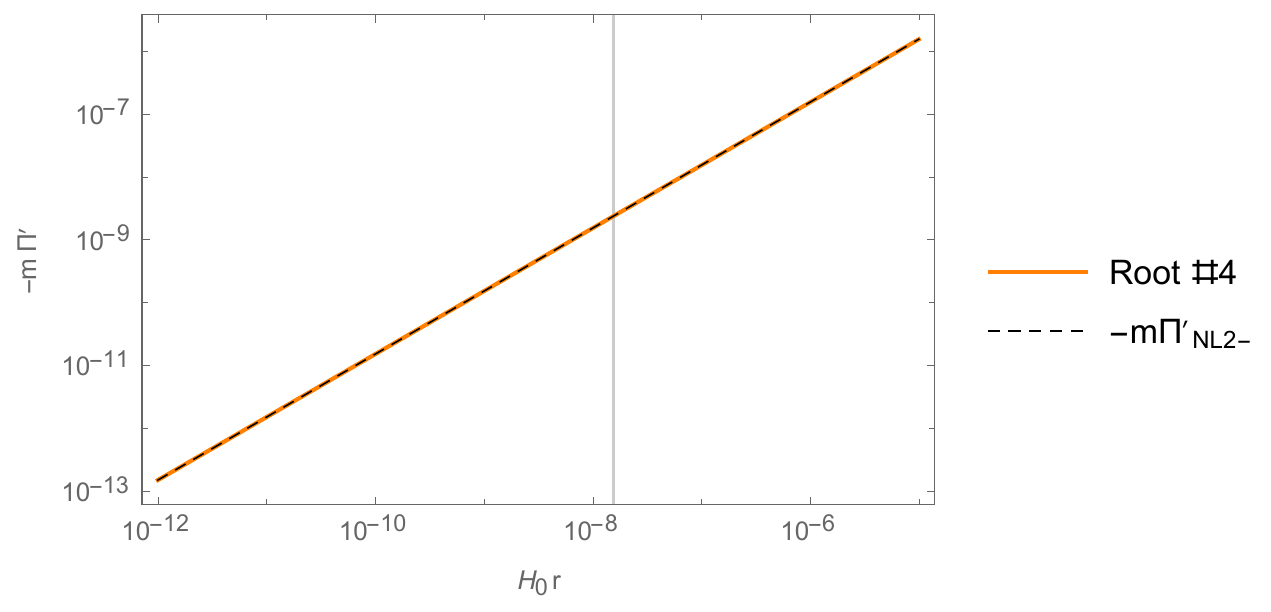}
	\caption{The root that matches the mass-independent non-linear behaviour 
	\eqref{eq:pi'NL2}. The vertical line shows the radius where this solution coincides with the linear solution \eqref{eq:pi'linear}.}
 	\label{fig:PLOTNL2}
\end{figure}

In Figure~\ref{fig:PLOTNL2} we show that the root \#4 matches $\Pi'_{NL2-}$, while $\Pi'_{NL2+}$ has a positive sign so is not presented. Again, the root that has the mass-independent behaviour at the non-linear regime fails to recover the linear solution.

\subsubsection{Neighbourhood of the singular point}

Finally, the third behaviour is triggered by the denominator in the last term of Eq.\eqref{eq:mastereq-formal}, which stems from the coefficient of the temporal St\"uckelberg perturbation in Eq.\eqref{eq:ST0integrated}. In our numerical result, the $\Pi'$ solution that matches the right linear behaviour becomes larger as $r$ decreases, thus potentially triggering a divergent behaviour. As a result, some the zeroes of the master equation approach the neighbourhood of the singular point. We define the value of $\Pi'$ at the singular point as
\begin{equation}
\mathcal{C}_1+\mathcal{C}_2\,\frac{\Pi_s'}{r} + \mathcal{C}_3\,\left(\frac{\Pi_s'}{r}\right)^2=0\,,
\label{eq:NL3}
\end{equation}
or,
\begin{equation}
\Pi'_{s\pm} = -r\,\frac{\mathcal{C}_2\pm\sqrt{\mathcal{C}_2^2-4\,\mathcal{C}_1\mathcal{C}_3}}{2\,\mathcal{C}_3}\,.
\label{eq:pi'singular}
\end{equation}
We then look for solutions that are in the neighbourhood of these values, $\Pi' = \Pi'_{s} + \delta \Pi'$. Expanding the master equation for small $r$ and small $\delta\Pi'$,
we get
\begin{equation}
\frac{G\,\delta M\,\left(\tilde{c}-1-\frac{\Pi_s'}{r}\right)\,\left(2\,\mathcal{C}_1+\mathcal{C}_2\frac{\Pi_s'}{r}\right)^2}{r^3\,a^3\,\left(\frac{\Pi_s'}{r}\right)^2}
-\frac{\tilde{c}+1+\frac{\Pi_s'}{r}}{2\,\tilde{c}^2\frac{(\delta\Pi')^2}{r^2}}\left[
\mathcal{C}_1(\tilde{c}-1)\,H + \dot{\mathcal{C}}_1\frac{\Pi_s'}{r}+\frac{\dot{\mathcal{C}}_2}{2}\,\left(\frac{\Pi_s'}{r}\right)^2+\frac{\dot{\mathcal{C}}_3}{3}\,\left(\frac{\Pi_s'}{r}\right)^3
\right]^2=0\,,
\end{equation}
where we have eliminated $\mathcal{C}_3$ by using \eqref{eq:NL3}.
If 
\begin{equation}
\frac{\tilde{c}+1+\frac{\Pi_s'}{r}}{\tilde{c}-1-\frac{\Pi_s'}{r}}>0\,,
\end{equation}
then there is a real $\delta\Pi'$ solution. In this case, we find
\begin{equation}
\delta\Pi' = \pm\sqrt{\frac{a^3r^5\left(\tilde{c}+1+\frac{\Pi_s'}{r}\right)}{2\,G\,\delta M\,\left(\tilde{c}-1-\frac{\Pi_s'}{r}\right)}}
\,\frac{\frac{\Pi_s'}{r}}{\left(2\,\mathcal{C}_1+\mathcal{C}_2\frac{\Pi_s'}{r}\right)\tilde{c}}\,\left[
\mathcal{C}_1(\tilde{c}-1)\,H + \dot{\mathcal{C}}_1\frac{\Pi_s'}{r}+\frac{\dot{\mathcal{C}}_2}{2}\,\left(\frac{\Pi_s'}{r}\right)^2+\frac{\dot{\mathcal{C}}_3}{3}\,\left(\frac{\Pi_s'}{r}\right)^3
\right]\,.
\end{equation}
We represent the four possible non-linear solutions of the third type as
\begin{equation}
\Pi_{NL3\pm\pm} = \Pi'_{s\pm} \pm |\delta \Pi'|\,,
\label{eq:pi'NL3}
\end{equation}
where the first sign in the subscript corresponds to the singular solution, while the second sign determines how $|\delta\Pi'|$ is introduced. Note that the dominant contribution is the singular solution which is linear in $r$ but the correction term goes as $r^{5/2}$. In our example, we find that all four of these estimates have the same sign as the linear solution.

\begin{figure}[h!]
	\centering
	\includegraphics[width=0.49\columnwidth]{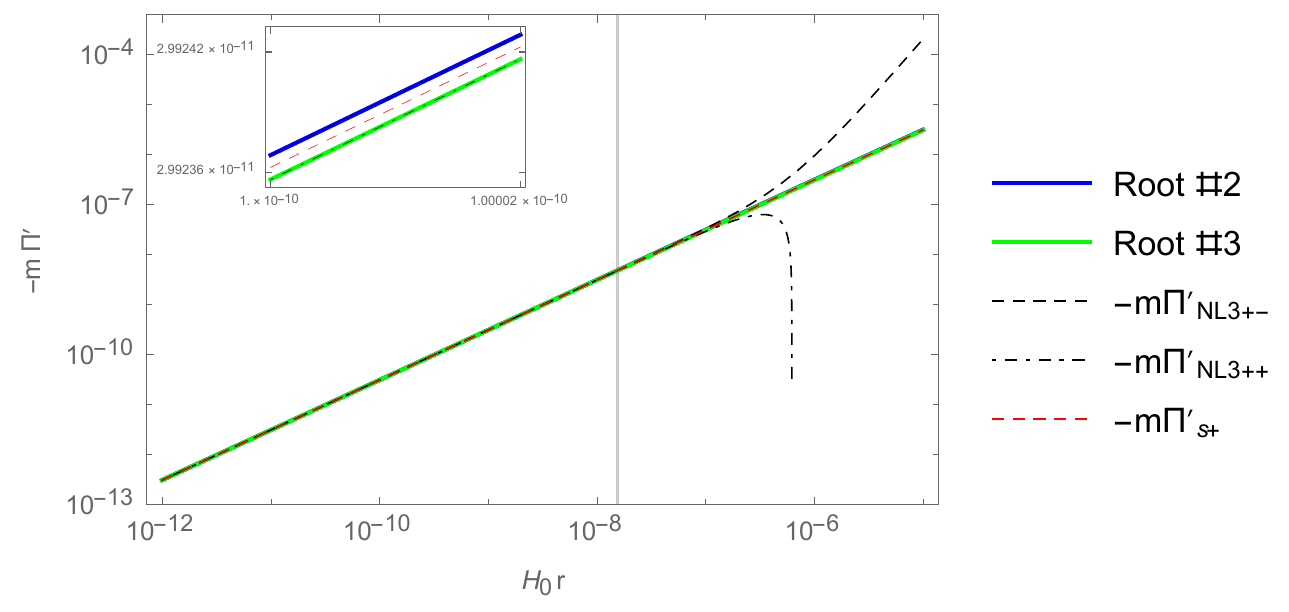}
	\includegraphics[width=0.49\columnwidth]{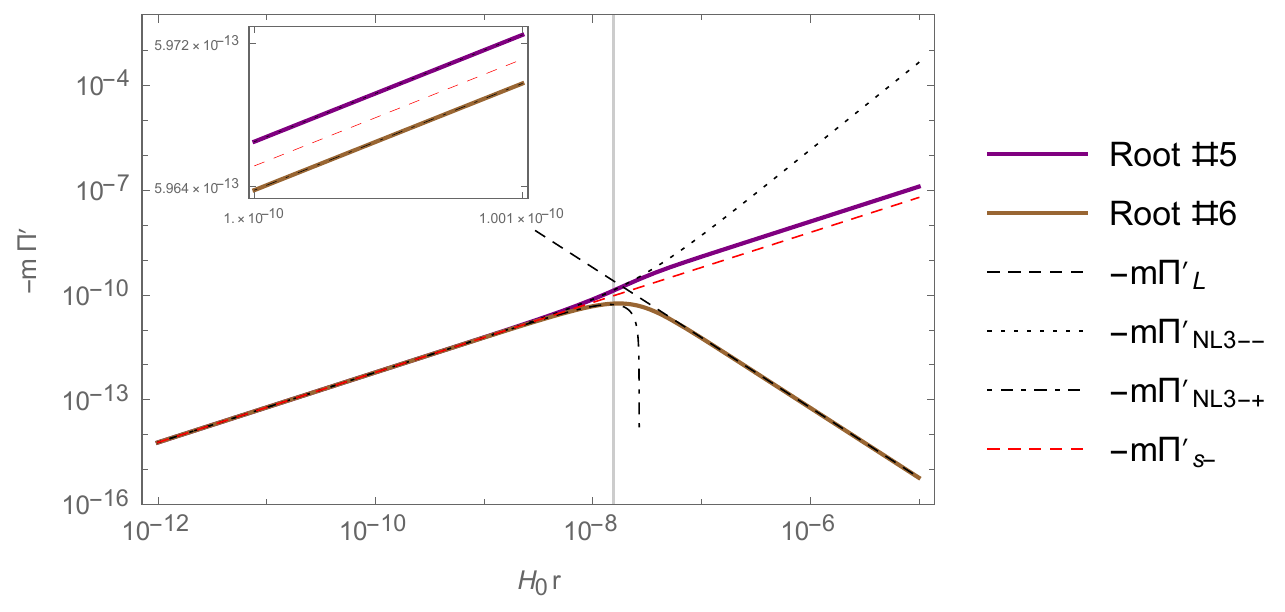}
	\caption{The roots that match the analytic estimates in the neighbourhood of the singular point. In the left (right) panel, we show the roots that match $\Pi'_{NL3+\pm}$ ($\Pi'_{NL3-\pm}$) solutions \eqref{eq:pi'NL3}. The vertical line shows the radius where the corresponding singular solution coincides with the linear solution \eqref{eq:pi'linear}.}
 	\label{fig:PLOTNL3}
\end{figure}

In Figure \ref{fig:PLOTNL3}, we show that the roots \#2 and \#3 coincide with the estimates $\Pi_{NL+\pm}$ while never intersecting the singular solution $\Pi_{s+}$, but failing to recover Eq.\eqref{eq:pi'linear} at large distances. In the right panel of Figure~\ref{fig:PLOTNL3} we also show that the roots \#5 and \#6 follow $\Pi_{NL-\pm}$ at short distances. In particular, the root \#6 is the solution that recovers the linear solution so we conclude that the associated non-linear behaviour is accurately described by the analytic solution $\Pi_{NL-+}$.

Finally, we discuss the Vainshtein radius where the linear solution matches the non-linear one. However, for $\Pi'_{NL3-+}$ solution that transitions into $\Pi'_{L}$, the solution crosses zero, failing to merge with the linear solution. This is due to the approximation of small $|\delta\Pi'|$ assumption we made while estimating this solution breaking down. We instead define the Vainshtein radius as the distance where the singular solution $\Pi'_{s-}$ matches the linear solution,
\begin{equation}
 \Pi'_{L} - \Pi'_{s-}\Big\vert_{r=r_V} = 0\,,
\label{eq:vainshtein-def}
\end{equation}
which is compatible with the small $\delta\Pi'$ approximation. Using our numerical results, we find,
\begin{equation}
\,r_V\simeq \left(\frac{G\,\delta M}{ m^2}\right)^{1/3}\,.
\end{equation}
Since $m \sim H_0$, this agrees with the expression of Vainshtein radius in the decoupling limit theories of massive gravity or galileon theories \cite{Babichev:2013usa}. 
For the sun, the Vainshtein radius is roughly given by $r_V \sim 100~{\rm pc}$. 

\begin{figure}[h!]
	\centering
	\includegraphics[width=0.7\columnwidth]{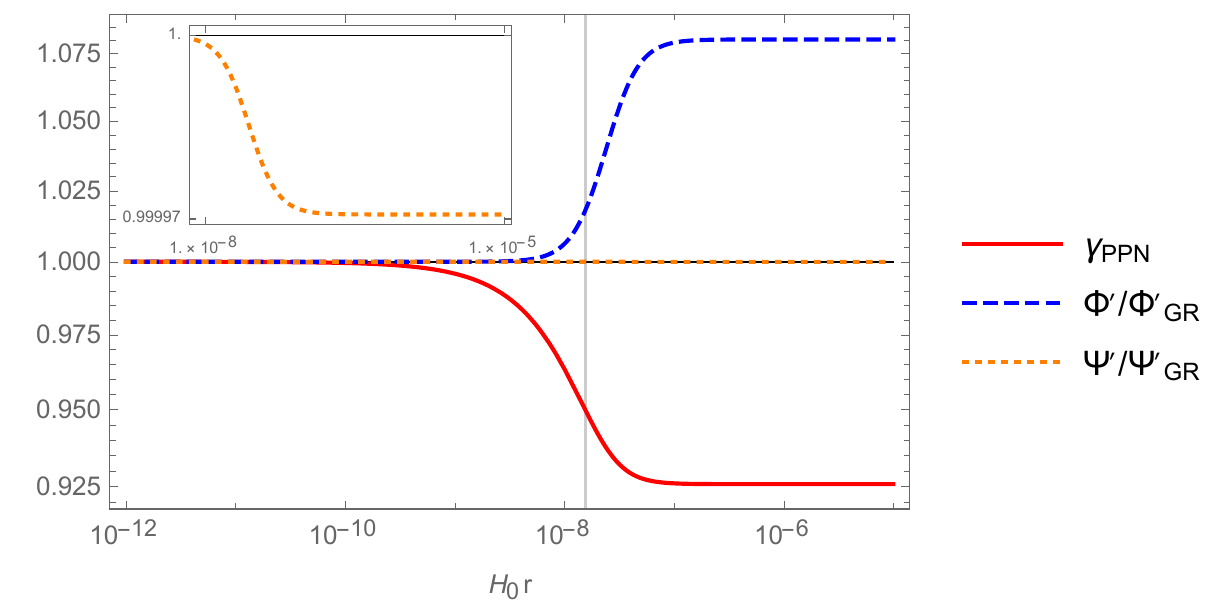}
	\caption{
		The PPN parameter $\gamma_{\rm PPN}=\Psi/\Phi$, and the values of $\Phi'$ and $\Psi'$ relative to their GR values. The vertical line shows the radius where this solution coincides with the linear solution \eqref{eq:pi'linear}.
		}
	\label{fig: PPN}
\end{figure}

Finally, 
we
discuss the effect on gravitational potentials. In Fig.~\ref{fig: PPN}, we show the parametrized post-Newtonian (PPN) parameter $\gamma_{\rm PPN}=\Psi/\Phi$, and the ratios $\Phi'/\Phi'_{\rm GR}$ and $\Psi'/\Psi'_{\rm GR}$ computed from \eqref{eq:solpsi'} and \eqref{eq:solphi'}, where $\Phi'_{\rm GR}$ and $\Psi'_{\rm GR}$ are gravitational potentials in GR. 
Inside the Vainshtein radius, all three quantities converge to their GR values while they deviate in the linear regime.
For instance, at $r=0.1$~pc, or $H_0\,r \sim 10^{-11}$, the deviation of the PPN parameter from unity is $\mathcal{O}(10^{-5})$.
The deviation of the PPN parameter well inside the Vainshtein radius is scaled as $r^3$ since the correction to the Newtonian contribution in $\Psi$ and $\Phi$ is proportional to $r^2$ for the nonlinear solution \eqref{eq:pi'NL3}, thus $|1-\gamma_{\rm PPN}| \sim {\cal O}(10^{-14})$ at the distance to Saturn, $r = 10 \,{\rm AU}$,  
which is consistent with the solar system tests $|1-\gamma_{\rm PPN}| < 2.3 \times 10^{-5}$ \cite{Will:2014kxa}. On the other hand, the Newtonian potential $\Phi$ tends to be larger outside the Vainshtein radius, meaning the additional attractive force enhances the effective gravitational constant.

\subsection{Linear approximation for $\delta\chi$}

In the previous subsection, we showed that the solution for $\Pi'$ that reduces to the correct linear behaviour is in the neighbourhood of a singular point at small $r$. This stems from the temporal St\"uckelberg equation \eqref{eq:ST0integrated}, where the coefficient of $\delta\chi$ becomes small. This brings the possibility that the assumption $\delta\chi \sim \mathcal{O}(\epsilon^2)$ might be inconsistent at short distances. We therefore need to check whether the linear solution for $\delta\chi'$ is still valid.

\begin{figure}[h!]
	\centering
	\includegraphics[width=0.7\columnwidth]{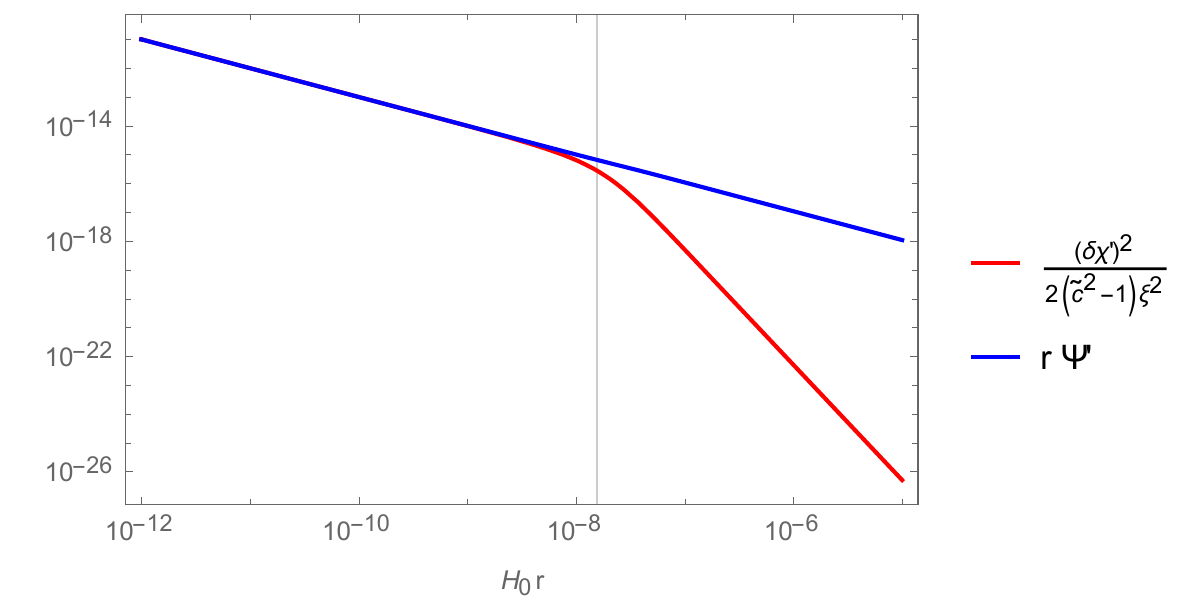}
	\caption{The comparison of $r\Psi'$ with our analytic estimate in \eqref{eq:dchiestimate}, using the numerical solutions for $\Psi'$ and $\delta\chi'$.}
 	\label{fig:checkestimate}
\end{figure}

We use the analytic solution $\Pi_{NL3--}$ in the linear solution for $\delta\chi'$ \eqref{eq:dchisol}. We then expand for small $r$. For the background evolution at hand, we have $\mathcal{C}_1 \sim \mathcal{O}(10^{-2})$, $\mathcal{C}_2\sim\mathcal{O}(1)$, $\mathcal{C}_3\sim \tilde{c} \sim \mathcal{O}(10)$, evaluated today. We can thus assume\footnote{This approximation might not be always true since we focused on only $Q=1$ case in the model \eqref{model}. Nonetheless we expect that the conclusion here does not significantly change. }
\begin{equation}
\vert\mathcal{C}_2 - \sqrt{\mathcal{C}_2^2 -4\,\mathcal{C}_1\mathcal{C}_3}\vert \ll 2\,\mathcal{C}_3(\tilde{c}\pm1)\,.
\end{equation}
As a result, we estimate $\delta\chi'$ at short distances as
\begin{equation}
\delta\chi' \simeq -\frac{\sqrt{2\,G\,\delta M\,(\tilde{c}^2-1)}\,\xi}{\sqrt{r}}\,.
\end{equation}
Similarly estimating the $\Psi'$ solution \eqref{eq:solpsi'}, we obtain
\begin{equation}
\Psi' \simeq \frac{G\,\delta M}{r^2}\,.
\end{equation}
We can thus relate $\delta \chi'$ to $\Psi'$ via
\begin{equation}
(\delta\chi')^2 \simeq 2\, (\tilde{c}^2-1)\xi^2 \,r\,\Psi'\,.
\label{eq:dchiestimate}
\end{equation}
Using the background values evaluated today, this reduces to $(\delta\chi')^2 \sim10^{3}\,r\,\Psi'$. This estimate is confirmed using numerical results in Figure \ref{fig:checkestimate}.

We are interested in the weak gravity regime, which requires  $\Psi \ll r\,\Psi'$. In this particular example due to the numerical factor, the linear approximation $\delta \chi' \ll 1$ breaks down at a distance larger than the Schwarzschild radius. Nonetheless, for the Sun, $r \Psi'$ is at most $10^{-5}$ and the linear approximation for $\delta \chi'$ is always valid. 

\section{Conclusions}
\label{sec6}
In this paper we studied the Vainshtein mechanism in Generalised Massive Gravity. GMG offers a possibility to explain the late-time accelerated expansion of the universe via the mass of the graviton. The theory admits open FLRW solutions on which 2 tensor, 2 vector and 1 scalar perturbations propagate \cite{Kenna-Allison:2019tbu}, unlike the cosmology in original dRGT theory \cite{Gumrukcuoglu:2011zh} where only the tensor modes survive at linear order. In GMG, the behaviour of linear perturbations are modified with respect to GR when the mass term dominates the expansion. Furthermore, the growth of structure is enhanced due to the propagation of the additional scalar mode and the tensor modes acquire a time-dependent mass \cite{Kenna-Allison:2020egn}.

The modification of the Newton's constant on linear (large) scales
indicates that the theory needs a screening mechanism to suppress the scalar mode on non-linear (small) scales to pass the Solar System constraints \cite{Will:2014kxa,deRham:2016nuf}. In this paper, we investigated the non-linear interactions of the scalar mode to identify the existence of the Vainshtein mechanism. Since the original dRGT model is inspired by the galileon theories, which accommodate the Vainshtein mechanism due to non-linearities of the second derivative of the scalar field, we expect that a similar mechanism exists in GMG. We first analysed cosmological perturbations using the quasi-static approximation. In the analysis we included second order interactions of the second derivative of the 
spatial component of the St\"uckelberg field $\Pi$. The structure of the non-linear terms in the Einstein equations are identical to those in the Horndeski theories, while the equation obtained by integrating out the time component of the St\"uckelberg field shows an interesting difference: it contains the third spatial derivative as well as the time derivative of $\Pi$. This is reminiscent of DHOST theories. However, contrary to these theories, the non-linear interactions in GMG do not truncate at finite order. This is because of the matrix square-root in the graviton potential. By expanding it, we found an infinite series of second derivative interactions of $\Pi$. The solution for the time component of the St\"uckelberg field also includes an infinite series of the second derivative of $\Pi$. 

In order to obtain interactions at an arbitary order, we considered spherically symmetric fluctuations around FLRW in a non-perturbative setup. Thanks to the symmetry, it was possible to compute the square-root matrix without making a perturbative expansion. In the weak field limit, we obtained the non-perturbative solution for the time component of the St\"uckelberg field. The solution includes the radial derivative of $\Pi$ in the denominator. We showed that, by expanding the solution, the perturbative solution was reproduced. We then obtained a master equation for the first radial derivative of $\Pi$. Since we integrated out the time component, the matter equation also has the radial derivative of $\Pi$ in the denominator. For a model considered in \cite{Kenna-Allison:2020egn}, we numerically solved the master equation to find solutions for metric perturbations and $\Pi$. We also found analytic solutions at small radius for which non-linear terms dominate over the linear contribution. We found that one of the non-linear solutions connects to the linear solution on large scales. The Vainshtein radius is given by $\,r_V \simeq \left(H_0^{-2} \,G\,\delta M\right)^{1/3}\,$ where the present-day Hubble constant appears because of our choice $m \sim H_0$. This is the same expression as in the Horndeski theories. In this particular example, the non-linear solution that connects to the linear solution is driven by the contribution from the temporal component of the St\"uckelberg field $\delta \chi$, which has $\Pi'$ in the denominator. At small scales, the denominator approaches zero and this is balanced by the non-linear coupling term between the gravitational potential and $\Pi'$. This is a new type of non-linear solution that does not exist in galileon-type theories. The new solution behaves as $\Pi' \propto r$ and is highly suppressed compared with the linear solution $\Pi_L' \propto 1/r^2$, realising the Vainshtein mechanism to suppress the modification of gravity below the Vainshtein radius. 

With this study, we showed that the cosmological model in the minimal GMG theory is also compatible with local tests of gravity. Although some of the features of the Vainshtein mechanism are similar to scalar-tensor theories, the distinct interaction between the metric and the scalar fields give rise to a new type of solution for the scalar fields in the non-linear regime. In particular, for the simple example given here, the PPN parameter $\gamma_{\rm PPN}$ is well within the current bounds.
Combined with the background cosmology, the evolution of scalar perturbations in the linear regime and the related evolution of the tensor mass \cite{Kenna-Allison:2020egn}, this model is a potentially falsifiable alternative to dark energy that can be differentiated from cosmological models in the scalar-tensor theory class.

A potential concern regarding the the cosmological model studied here is the behaviour at early times. In the early universe, the rapid time variation in $\alpha_2$ is effectively suppressed and the cosmology approaches to the self-accelerating branch of the original dRGT theory. In dRGT, this branch is infinitely strongly coupled due to an exact vanishing of $\CC_1$. In contrast, this quantity is generically non-zero in our model, although it decreases as we go to the past. On the other hand, the minimal model is a single point in the theory space; the theory class that GMG belongs to \cite{Gumrukcuoglu:2020utx} admits six arbitrary functions which can in principle be arranged to give slow variation such that the strong coupling remains finite without invalidating the cosmological solution. A future study of non-linear perturbations in a more general setup can allow us to determine how the strong coupling scale  associated with the early universe solutions depend on the variation of the mass functions.

\acknowledgments 
KK is supported by the European Research Council under the European Union's Horizon 2020 programme (grant agreement No.646702 ``CosTesGrav") and the UK STFC grant ST/S000550/1. AEG is supported by a Dennis Sciama Fellowship at the University of Portsmouth.

\appendix
\section{Equations of motion for non-linear cosmological perturbations}
\label{app:cosmoeq}
In this Appendix, we present all equations for non-linear cosmological perturbations up to second order. 

\subsection{Einstein's equations}
First and second order Einstein's equations are as follows. The time-time component is:
\begin{align}
	\delta^{(1)}\mathcal{E}^0_0 =& -\frac{\delta\rho}{M_p^2}-m^2\,\CC_1\,\partial^2\Pi + \frac{2}{a^2}\partial^2\Psi\,,
	\nonumber\\
	\delta^{(2)}\mathcal{E}^0_0= & -\frac{m^2\CC_2}{4} \left[(\partial^2\Pi)^2-(\partial_i\partial_j\Pi)^2
	\right]\,.
\end{align}
For the spatial part, the trace part  $\mathcal{E}^{\rm tr} \equiv \frac{\delta^{j}_i}{3}\,\mathcal{E}^i_j$ is obtained as
\begin{align}
	\delta^{(1)} \mathcal{E}^{\rm tr} =& \frac{m^2\,[2\,\CC_1+(\ct-1)\,\CC_2]}{3}\,\partial^2\Pi +\frac{2}{3\,a^2}\partial^2(\Phi-\Psi)\,,
	\nonumber\\
	\delta^{(2)} \mathcal{E}^{\rm tr} =& \frac{m^2[\CC_2+2\,(\ct-1)\,\CC_3]}{12}\,\left[(\partial^2\Pi)^2-(\partial_i\partial_j\Pi)^2\right]\,,
\end{align}
while the traceless part, defined as $\mathcal{E}^{{\rm trless}\,i}_{\;\;\;\;\;\;j} \equiv \mathcal{E}^i_j -\delta^i_j\,\mathcal{E}^{\rm tr}$, is 
\begin{align}
	\delta^{(1)} \mathcal{E}^{{\rm trless}\,i}_{\;\;\;\;\;\;j} =& -\frac{1}{a^2}\left(\partial^i\partial_j-\frac{\delta^{i}_j}{3}\partial^2\right)\left(\Phi-\Psi+\frac{m^2\,a^2\,[2\,\CC_1+(\ct-1)\CC_2]}{2}\,\Pi\right)\,,
	\nonumber\\
	\delta^{(2)} \mathcal{E}^{{\rm trless}\,i}_{\;\;\;\;\;\;j} =&
	\frac{m^2[\CC_2+2\,(\ct-1)\,\CC_3]}{2}\,\left[\left(\partial^i\partial^k\Pi\,\partial_k\partial_j\Pi - \frac{\delta^i_j}{3} (\partial_k\partial_l\Pi)^2\right)
	-\left(
	\partial^i\partial_j\Pi\,\partial^2\Pi - \frac{\delta^i_j}{3} (\partial^2\Pi)^2
	\right)
	\right]\,.
\end{align}
We can define a scalar quantity from the traceless equations via
\begin{equation}
	\partial_j \mathcal{E}^{{\rm trless}} \equiv \partial_i\mathcal{E}^{{\rm trless}\,i}_{\;\;\;\;\;\;j}\,,
\end{equation}
which is given by (truncated at quadratic order in perturbations)
\begin{equation}
	\mathcal{E}^{{\rm trless}} =  -\frac{2}{3\,a^2}\,\partial^2 \left( \Phi-\Psi+\frac{m^2a^2\,[2\,\CC_1+\CC_2(\ct-1)]}{2}\,\Pi\right) -\frac{m^2[\CC_2+2\,\CC_3(\ct-1)]}{12}\,[(\partial^2\Pi)^2-(\partial_i\partial_j\Pi)^2]\,.
\end{equation}

\subsection{St\"uckelberg equations}
\label{sec:stuckelbergeq}
\subsubsection{Temporal component}
We find the St\"uckelberg equations of motion as follows:
\begin{align}
	\delta^{(1)}\mathcal{E}^{\rm St}_0=& \frac{1}{a^2}\partial^2\left[
	-\frac{\ct^2\,\CC_1}{\ct+1}\,\left(\Pi^0+\frac{a^2}{{\tilde c}}\,\dot{\Pi}
	\right)
	-a^2\,\left(\dot{\CC}_1+H\,[\CC_1-(\ct-1)\,\CC_2]\right)\,\Pi
	\right]\,,\nonumber\\
	\delta^{(2)}\mathcal{E}^{\rm St}_0=&\frac{1}{a}
	\partial_i\Bigg\{\frac{\ct^2}{2\,a\,(\ct+1)}\,\partial^j\left(\Pi^0+\frac{a^2}{{\tilde c}}\dot\Pi\right)\left[\left(\CC_2+\frac{2\,\CC_1}{\ct+1}\right)\partial^i\partial_j\Pi - \CC_2\,\delta^i_j\,\partial^2\Pi\right]
    \nonumber\\
	&\qquad\quad+\frac{a\,\left(\dot{\CC}_2+2\,H\,[\CC_2-(\ct-1)\,\CC_3]\right)}{4}\,\partial^j\Pi\left( \partial^i\partial_j\Pi- \delta^i_j\,\partial^2\Pi\right)
	\Bigg\}\,.
	\label{eq:stucktempeqs}
\end{align}
We notice that the linear and quadratic terms can be written as a divergence. Defining
\begin{equation}
	\frac{1}{a}\partial_i \delta\mathcal{E}^{{\rm St}\,i}_0 = \delta^{(1)}\mathcal{E}^{\rm St}_0+\delta^{(2)}\mathcal{E}^{\rm St}_0
\end{equation}
we can integrate the equation once to find
\begin{align}
	\delta\mathcal{E}^{{\rm St}\,i}_0 =&
	-\frac{\ct^2}{2\,a\,(\ct+1)}\,\partial^j\left(\Pi^0+\frac{a^2}{{\tilde c}}\dot\Pi\right)
	\left[
	2\,\delta^i_j\,\CC_1-
	\left(\CC_2+\frac{2\,\CC_1}{\ct+1}\right)\partial^i\partial_j\Pi + \CC_2\,\delta^i_j\,\partial^2\Pi\right]
	\nonumber\\
	& +a\,H\,\partial^j\Pi\left[ \left(-\CC_1+(\ct-1)\,\CC_2-\frac{\dot{\CC}_1}{H}\right)\delta^i_j\,
	+ \left(
	\frac{\CC_2-(\ct-1)\,\CC_3}{2} + \frac{\dot{\CC}_2}{4\,H}
	\right)\,\left( \partial^i\partial_j\Pi- \delta^i_j\,\partial^2\Pi\right)\right]\,.
	\label{eq:stuckelbergT-perturbative}
\end{align}
We can solve this equation perturbatively for $\Pi^0$. Formally we have
\begin{equation}
	\mathcal{M}_1 \,\partial \left(\Pi^0+\frac{a^2}{\ct}\dot\Pi\right)= \mathcal{M}_2 \,\partial\Pi \,,
\end{equation}
which is solved by $\partial \left(\Pi^0+\frac{a^2}{\ct}\dot\Pi\right)= M_1^{-1} M_2 \partial\Pi$. Expanding these matrices perturbatively, we find 
\begin{align}
	\partial_i \Pi^0 =& -\frac{a^2}{\ct}\,\partial_i\dot\Pi 
	-\frac{a^2(\ct+1)}{4\,\ct^2\,\CC_1^2}\left[4\,\CC_1\left(\dot{\CC}_1+H\,[\CC_1-(\ct-1)\,\CC_2]\right)
	-[2\,\CC_2\,\dot{\CC}_1-\CC_1\dot{\CC}_2-2\,H\,(\ct-1)\,(\CC_2^2-\CC_1\CC_3)]
	\,\partial^2\Pi\right]\partial_i\Pi 
	\nonumber\\
	&
 	-\frac{a^2}{4\,\ct^2\,\CC_1^2}\,
 	\left(
 	2\,\CC_2\,\dot{\CC}_1(\ct+1)+\CC_1[4\,\dot{\CC}_1-(\ct+1)\,\dot{\CC}_2]
 	+H\,[4\,\CC_1^2-2\,\CC_2^2(\ct^2-1)+2\,\CC_1(\ct-1)(-2\,\CC_2+\CC_3(\ct+1))]
 	\right)
\nonumber\\
&\qquad\qquad\qquad \times \partial_i\partial^j\Pi\,\partial_j\Pi\,.
	\label{eq:pi0sol}
\end{align}
\subsubsection{Spatial component}
Spatial part of the St\"uckelberg equation at linear order is given by
\begin{align}
	\delta^{(1)}\mathcal{E}^{\rm St}_i = & \partial_i\Bigg[
	\CC_1\,\Phi-
	\,[2\,\CC_1+(\ct-1)\,\CC_2]\,\Psi - \frac{H\,[2\,(\ct+2)\,\CC_1+(\ct^2-1)\,\CC_2]}{\ct+1}\,\Pi^0 
	\nonumber\\
	&\qquad\qquad\qquad+ \frac{(\dot{\CC}_1+5\,H\,\CC_1)(\ct+1)-\CC_1\,\dot{\ct}}{(\ct+1)^2}
	\left(\Pi^0-a^2\dot\Pi\right)-  \frac{\ct\,\CC_1}{\ct+1}\,\left(\dot\Pi^0+\frac{a^2}{{\tilde c}}\,\ddot\Pi\right)
	\Bigg]\,.
	\label{eq:stuckilin}
\end{align}
%
%
At quadratic order, we have 
\begin{align}
	\delta^{(2)}\mathcal{E}^{\rm St}_i= & - \frac{\ct^2}{4\,(\ct+1)^2} \,\left(
\frac{1}{a^2}\left[2\,(2\,\ct+1)\,\CC_1-(\ct+1)\CC_2\right]\partial_i\Pi^0 
+ \left[2\,\CC_1+(\ct+1)\,\CC_2\right]\partial_i\dot\Pi\right)\partial^2\Pi^0
	\nonumber\\
	&-\frac{\ct}{4\,(\ct+1)^2}\left(\left[2\,(\ct-2)\,\CC_1+\ct\,(\ct+1)\,\CC_2\right]\partial^j\Pi^0 -\frac{a^2}{\ct} \left[6\,\CC_1+(\ct+1)\,\CC_2\right]\partial^j\dot\Pi\right)\,\partial_i\partial_j\dot\Pi\nonumber\\
	&
	-\frac{\ct^2[2\,\CC_1-(\ct+1)\,\CC_2]}{4\,(\ct+1)^2}\,\left(\partial_i\Pi^0-\frac{a^2}{r^2}\partial_i\dot\Pi\right)\,\partial^2\dot\Pi
	\nonumber\\
	&+\frac{\ct}{4\,(\ct+1)^2}\left(\frac{\ct}{a^2}\left[2\,\CC_1-(\ct+1)\,\CC_2\right]\,\partial^j\Pi^0+\left[2\,(\ct+2)\,\CC_1+\ct\,(\ct+1)\,\CC_2\right]\partial^j\dot\Pi\right) \partial_i\partial_j\Pi^0
	\nonumber\\
	&
	-\frac12\,\left(2\,\dot{\CC}_1 + H\,[2\,\CC_1+ 2\,(\ct-1)\,\CC_3-(2\,\ct-3)\,\CC_2] -\frac{\dot\CC_2+3\,H\,\CC_2}{\ct+1}+\frac{\dot\ct\,\CC_2}{(\ct+1)^2}\right)\,\partial^2\Pi\,\partial_i\Pi^0
	\nonumber\\
	&
	-\frac{a^2}{2(\ct+1)}\left(\dot\CC_2+5\,H\,\CC_2 - \frac{\dot\ct\,\CC_2}{\ct+1}\right) \,\partial^2\Pi\,\partial_i\dot\Pi
	\nonumber\\
	&
	-\frac{1}{2}\Bigg(
	H\,\left[4\,\CC_1+(2\,\ct-3)\,\CC_2-2\,(\ct-1)\,\CC_3\right]
	-\frac{4\,\dot\CC_1-\dot\CC_2+3\,H\,(4\,\CC_1-\CC_2)}{\ct+1}
	\nonumber\\
	&\qquad\qquad\qquad\qquad\qquad\qquad\qquad\qquad+\frac{2\,\dot\CC_1+6\,H\,\CC_1+\dot\ct\,(4\,\CC_1-\CC_2)}{(\ct+1)^2}
	-\frac{4\,\dot\ct\,\CC_1}{(\ct+1)^3}
	\Bigg)\partial^j\Pi^0\partial_i\partial_j\Pi
	\nonumber\\
	&
	-\frac{a^2}{2\,(\ct+1)}\,\left(
	2\,\dot\CC_1-\dot\CC_2+5\,H\,(2\,\CC_1-\CC_2)-\frac{2\,\dot\CC_1-\dot\ct\,\CC_2+2\,(5\,H+\dot\ct)\,\CC_1}{\ct+1}
	+\frac{4\,\dot\ct\,\CC_1}{(\ct+1)^2}\right)\partial^j\dot\Pi\,\partial_i\partial_j\Pi
	\nonumber\\
	&
	-\frac{\ct}{2\,(\ct+1)^2}\,\left(\partial_j\dot\Pi^0 + \frac{a^2}{r}\,\partial_j\ddot{\Pi}\right)\,\left[
	(\ct+1)\,\CC_2\,\delta^j_i\,\partial^2\Pi - [\CC_2-\ct\,(2\,\CC_1-\CC_2)]\,\partial_i\partial^j\Pi
	\right]
	\nonumber\\
	&
	+\frac{\CC_2}{2}\,\partial^2\Pi\,\partial_i\Phi 
	+\left(\CC_1-\frac{\CC_2}{2}\right)\,\partial_i\partial_j\Pi\,\partial^j\Phi
	-\left(\frac{\CC_2}{2}+(\ct-1)\,\CC_3\right)\,\partial^2\Pi\,\partial_i\Psi
    \nonumber\\
	&
	-\left[2\,\CC_1+\left(\ct-\frac32\right)\,\CC_2-(\ct-1)\,\CC_3\right]\,\partial_i\partial^j\Pi\,\partial_j\Psi
	\,.
	\label{eq:stuckiquad}
\end{align}

One way to simplify these expressions is to use the solution for $\Pi^0$ obtained in Eq.\eqref{eq:pi0sol}. The linear and quadratic parts reduce to

\begin{align}
	\delta^{(1)}\mathcal{E}^{\rm St}_i\Big\vert_{\rm reduced} =& 
	\partial_i\left(
	\CC_1\,\Phi-[2\,\CC_1+\CC_2(\ct-1)]\,\Psi +\A_1\,a^2H^2\,\Pi
	\right)\,,
	\nonumber\\
	\delta^{(2)}\mathcal{E}^{\rm St}_i\Big\vert_{\rm reduced} =&
		\partial_j\Psi\,
	\left[\left(\frac{\CC_2}{2}+\CC_3(\ct-1)\right)\left(\partial_i\partial^j\Pi-\delta_i^j\partial^2\Pi\right)-[2\,\CC_1+\CC_2(\ct-1)]\,\partial_i\partial^j\Pi\right]
	\nonumber\\
	&-\partial_j\Phi\,
	\left[\frac{\CC_2}{2}\left(\partial_i\partial^j\Pi-\delta_i^j\partial^2\Pi\right)-\CC_1\,\partial_i\partial^j\Pi\right]+a^2\,H^2\,\partial_j\Pi\,
	\left[
	\A_2 \left(\partial_i\partial^j\Pi-\delta_i^j\partial^2\Pi\right)+\A_1 \partial_i\partial^j\Pi
	\right]
    \nonumber\\
	&+\frac{a^2H}{4\,\tilde{c}}
	\left(-2\,\CC_1+(\CC_2-2\,\CC_3)(\ct-1)+\frac{\CC_2^2(\ct^2-1)}{\CC_1} 
	+\frac{\CC_1(-2\,\dot{\CC}_1+\dot{\CC}_2)-\CC_2(\ct+1)\dot{\CC}_1}{H\,\CC_1}
\right)
\nonumber\\
&\qquad\qquad\qquad \qquad\qquad\qquad \qquad\qquad\qquad \qquad\qquad\qquad \times
\partial_j\left(\partial_i\Pi\,\partial^j\dot\Pi-\partial_i\dot\Pi\,\partial^j\Pi\right)\,,
\end{align}
where we defined
\begin{align}
 \A_1 \equiv & \frac{\ddot\CC_1}{H^2\,\ct} -\frac{(\ct+1)\,\dot\CC_1^2}{H^2\,\ct^2\,\CC_1} + \frac{\dot\CC_1}{H\,\ct^2\,\CC_1}\,\left[\CC_1\,\left(4\,\ct-2-\frac{\dot\ct}{H}\right)+2\,\CC_2(\ct^2-1)\right]
 -\frac{(\ct-1)\,\dot\CC_2}{H\,\ct}-\frac{\dot\ct\,(\CC_1+\CC_2)}{H\,\ct^2}
 \nonumber\\
 & 
 +\frac{\CC_1-(\ct-1)\,\CC_2}{2\,\ct^2\,\CC_1}\,\left[
 (8\,\ct-2)\,\CC_1+2\,(\ct^2-1)\,\CC_2+\frac{m^2\,\ct\,(\ct-1)\,\CC_1^2}{H^2} - \frac{\ct\,\CC_1\,\rho}{M_p^2H^2}
 \right]\,,
 \nonumber\\
 \A_2 \equiv &
 -\frac{\ddot\CC_2}{4\,H^2\,\ct} 
 - \frac{[2\,\CC_1+3\,(\ct+1)\,\CC_2]\,\dot\CC_1^2}{4\,H^2\,\ct^2\,\CC_1^2}
 + \frac{3\,(\ct+1)\,\dot\CC_1\,\dot\CC_2}{4\,H^2\,\ct^2\,\CC_1} 
 -\frac{2\,\CC_1^2-3\,(\ct^2-1)\,\CC_2^2-(\ct-1)\,\CC_1 \,[2\,\CC_2-3\,(\ct+1)\,\CC_3]}{2\,H\,\ct^2\,\CC_1^2}\,\dot\CC_1
 \nonumber\\
&
 - \left(\frac{(4\,\ct-3)\,\CC_1+3\,(\ct^2-1)\,\CC_2}{\CC_1}-\frac{\dot\ct}{H}\right)\,\frac{\dot\CC_2}{4\,H\,\ct^2}
 +\frac{(\ct-1)\,\dot\CC_3}{2\,H\,\ct} 
 +\frac{\dot\ct\,(\CC_2+\CC_3)}{2\,H\,\ct^2}
 -\frac{m^2\,(\ct-1)\,\CC_1\,[\CC_2-(\ct-1)\,\CC_3]}{4\,H^2\,\ct} 
 \nonumber\\
 &
 +\frac{[\CC_2-(\ct-1)\,\CC_3]\,\rho}{4\,M_p^2\,H^2\,\ct}
 -\frac{1}{\ct^2}\,\left(
 \frac{\CC_1}{2}+\CC_2+\frac{(\ct-1)\,(3\,\CC_2+2\,\CC_3)}{4}+\frac{(\ct-1)^2(\CC_2^2-2\,\CC_1\,\CC_3)\,[2\,\CC_1+3\,(\ct+1)\,\CC_2]}{4\,\CC_1^2}
 \right)\,.
 \label{eq:defA1A2}
\end{align}

\section{Master equation}
\label{app:mastereq}
In this Appendix, we present the explicit form of Eq.\eqref{eq:mastereq-formal}:
\begin{align}
\frac{\tilde{c}^2 \mathcal{E}^{\rm St}_r }{r\,a^2(1+\Pi'')}
=& 
\left(\mathcal{C}_1+\mathcal{C}_2 \frac{\Pi'}{r}+\mathcal{C}_3 \left(\frac{\Pi'}{r}\right)^2\right)\frac{\tilde{c}^2\Phi'}{a^2r}
-\left(2\,\mathcal{C}_1+\mathcal{C}_2(\tilde{c}-1)+
[\mathcal{C}_2+2\,\mathcal{C}_3(\tilde{c}-1)]\frac{\Pi'}{r}
\right) \frac{\tilde{c}^2\Psi'}{a^2r} 
\nonumber\\
&+
\frac{\left(\mathcal{C}_2+2\,\mathcal{C}_3\frac{\Pi'}{r}\right)\left(\tilde{c}+1+\frac{\Pi'}{r}\right)\left(\frac{\dot{\mathcal{C}}_1\Pi'}{r} + \frac{\dot{\mathcal{C}}_2}{2}\left(\frac{\Pi'}{r}\right)^2+\frac{\dot{\mathcal{C}}_3}{3}\,\left(\frac{\Pi'}{r}\right)^3+\frac{\dot{L}}{3}\right)^2}{2\,\left(\mathcal{C}_1+\mathcal{C}_2\,\frac{\Pi'}{r}+\mathcal{C}_3 \left(\frac{\Pi'}{r}\right)^2\right)^2
}
\nonumber\\
&-
\frac{\left(\frac{\dot{\mathcal{C}}_1\Pi'}{r} + \frac{\dot{\mathcal{C}}_2}{2}\left(\frac{\Pi'}{r}\right)^2+\frac{\dot{\mathcal{C}}_3}{3}\,\left(\frac{\Pi'}{r}\right)^3+\frac{\dot{L}}{3}\right)
\left[
\left(\tilde{c}+1-\frac{\Pi'}{r}\right)\dot{\mathcal{C}}_1+(\tilde{c}+1)\dot{\mathcal{C}}_2\frac{\Pi'}{r}+\left(\tilde{c}+1+\frac{\Pi'}{3\,r}\right)\dot{\mathcal{C}}_3\left(\frac{\Pi'}{r}\right)^2-\frac{2}{3}\,\dot{L}\right]
}{\mathcal{C}_1+\mathcal{C}_2\,\frac{\Pi'}{r}+\mathcal{C}_3 \left(\frac{\Pi'}{r}\right)^2}
\nonumber\\
&
+\tilde{c}\,\left(\frac{\ddot{\mathcal{C}}_1\Pi'}{r} + \frac{\ddot{\mathcal{C}}_2}{2}\left(\frac{\Pi'}{r}\right)^2+\frac{\ddot{\mathcal{C}}_3}{3}\,\left(\frac{\Pi'}{r}\right)^3+\frac{\ddot{L}}{3}\right)
-\dot{\tilde{c}}\left(\frac{\dot{\mathcal{C}}_1\Pi'}{r} + \frac{\dot{\mathcal{C}}_2}{2}\left(\frac{\Pi'}{r}\right)^2+\frac{\dot{\mathcal{C}}_3}{3}\,\left(\frac{\Pi'}{r}\right)^3+\frac{\dot{L}}{3}\right)
\nonumber\\
&
+\left(1+\frac{\Pi'}{r}\right)\,H\left[
\left(\tilde{c}-1+\frac{2\,\Pi'}{r}\right)\dot{\mathcal{C}}_1
+\left(\tilde{c}-1+\frac{\Pi'}{2\,r}\right)\dot{\mathcal{C}}_2\frac{\Pi'}{r}
+(\tilde{c}-1)\dot{\mathcal{C}}_3\left(\frac{\Pi'}{r}\right)^2+\dot{L}
\right]
\nonumber\\
&
-\left(\mathcal{C}_1+\mathcal{C}_2\frac{\Pi'}{r}+\mathcal{C}_3\left(\frac{\Pi'}{r}\right)^2\right)\left[
\left(\tilde{c}-1-\frac{\Pi'}{r}\right)\tilde{c}\,\dot{H} + \left(1+\frac{\Pi'}{r}\right)\dot{\tilde{c}}\,H
\right]
\nonumber\\
&
-\left(\tilde{c}-1-\frac{\Pi'}{r}\right)\,H^2
\Bigg[
\left(2\,\tilde{c}+1+\frac{\Pi'}{r}\right)\,\mathcal{C}_1
+\frac12\,\left(\tilde{c}^2-1+\frac{4\,\tilde{c}\,\Pi'}{r}+\left(\frac{\Pi'}{r}\right)^2\right)\mathcal{C}_2
\nonumber\\
&\qquad\qquad\qquad\qquad\qquad\qquad\qquad\qquad\qquad\qquad\qquad\qquad+\,\left(\tilde{c}^2-1+\frac{(2\,\tilde{c}-1)\Pi'}{r}\right)\frac{\mathcal{C}_3\Pi'}{r}
\Bigg]\,.
\label{eq:mastereq-explicit}
 \end{align}

Using the solutions for the Newtonian potentials derived in Eqs.\eqref{eq:solpsi'}, \eqref{eq:solphi'}, we find
\begin{align}
\frac{\tilde{c}^2 \mathcal{E}^{\rm St}_r }{r\,a^2(1+\Pi'')}
=& 
-\tilde{c}^2\left(
\frac{G\,\delta M}{a^3r^3}+\frac{-3\,H^2+m^2L+\rho/M_p^2}{6}\right)
\left[
\mathcal{C}_1+(\tilde{c}-1)\mathcal{C}_2+\left(2(\tilde{c}-1)-\frac{\Pi'}{r}\right)\mathcal{C}_3\frac{\Pi'}{r}
\right]
\nonumber\\
&
+\frac{m^2\tilde{c}^2\Pi'}{12\,r}\Bigg[
-6\,\mathcal{C}_1[3\,\mathcal{C}_1+2\,\mathcal{C}_2(\tilde{c}-1)]
-9\,[2\,\mathcal{C}_1\mathcal{C}_2+2\,\mathcal{C}_1\mathcal{C}_3(\tilde{c}-1)+\mathcal{C}_2^2(\tilde{c}-1)]\frac{\Pi'}{r} 
\nonumber\\
&\qquad\qquad\quad- [3\,\mathcal{C}_2^2+8\,\mathcal{C}_1\,\mathcal{C}_3+20\,\mathcal{C}_2\mathcal{C}_3(\tilde{c}-1)]\left(\frac{\Pi'}{r}\right)^2
-10\,\mathcal{C}_3^2(\tilde{c}-1)\left(\frac{\Pi'}{r}\right)^3+2\,\mathcal{C}_3^2\left(\frac{\Pi'}{r}\right)^4
\Bigg]
\nonumber\\
&+
\frac{\left(\mathcal{C}_2+2\,\mathcal{C}_3\frac{\Pi'}{r}\right)\left(\tilde{c}+1+\frac{\Pi'}{r}\right)\left(\frac{\dot{\mathcal{C}}_1\Pi'}{r} + \frac{\dot{\mathcal{C}}_2}{2}\left(\frac{\Pi'}{r}\right)^2+\frac{\dot{\mathcal{C}}_3}{3}\,\left(\frac{\Pi'}{r}\right)^3+\frac{\dot{L}}{3}\right)^2}{2\,\left(\mathcal{C}_1+\mathcal{C}_2\,\frac{\Pi'}{r}+\mathcal{C}_3 \left(\frac{\Pi'}{r}\right)^2\right)^2
}
\nonumber\\
&-
\frac{\left(\frac{\dot{\mathcal{C}}_1\Pi'}{r} + \frac{\dot{\mathcal{C}}_2}{2}\left(\frac{\Pi'}{r}\right)^2+\frac{\dot{\mathcal{C}}_3}{3}\,\left(\frac{\Pi'}{r}\right)^3+\frac{\dot{L}}{3}\right)
\left[
\left(\tilde{c}+1-\frac{\Pi'}{r}\right)\dot{\mathcal{C}}_1+(\tilde{c}+1)\dot{\mathcal{C}}_2\frac{\Pi'}{r}+\left(\tilde{c}+1+\frac{\Pi'}{3\,r}\right)\dot{\mathcal{C}}_3\left(\frac{\Pi'}{r}\right)^2-\frac{2}{3}\,\dot{L}\right]
}{\mathcal{C}_1+\mathcal{C}_2\,\frac{\Pi'}{r}+\mathcal{C}_3 \left(\frac{\Pi'}{r}\right)^2}
\nonumber\\
&
+\tilde{c}\,\left(\frac{\ddot{\mathcal{C}}_1\Pi'}{r} + \frac{\ddot{\mathcal{C}}_2}{2}\left(\frac{\Pi'}{r}\right)^2+\frac{\ddot{\mathcal{C}}_3}{3}\,\left(\frac{\Pi'}{r}\right)^3+\frac{\ddot{L}}{3}\right)
-\dot{\tilde{c}}\left(\frac{\dot{\mathcal{C}}_1\Pi'}{r} + \frac{\dot{\mathcal{C}}_2}{2}\left(\frac{\Pi'}{r}\right)^2+\frac{\dot{\mathcal{C}}_3}{3}\,\left(\frac{\Pi'}{r}\right)^3+\frac{\dot{L}}{3}\right)
\nonumber\\
&
+\left(1+\frac{\Pi'}{r}\right)\,H\left[
\left(\tilde{c}-1+\frac{2\,\Pi'}{r}\right)\dot{\mathcal{C}}_1
+\left(\tilde{c}-1+\frac{\Pi'}{2\,r}\right)\dot{\mathcal{C}}_2\frac{\Pi'}{r}
+(\tilde{c}-1)\dot{\mathcal{C}}_3\left(\frac{\Pi'}{r}\right)^2+\dot{L}
\right]
\nonumber\\
&
-\left(\mathcal{C}_1+\mathcal{C}_2\frac{\Pi'}{r}+\mathcal{C}_3\left(\frac{\Pi'}{r}\right)^2\right)\left[
\left(\tilde{c}-1-\frac{\Pi'}{r}\right)\tilde{c}\,\dot{H} + \left(1+\frac{\Pi'}{r}\right)\dot{\tilde{c}}\,H
\right]
\nonumber\\
&
-\left(\tilde{c}-1-\frac{\Pi'}{r}\right)\,H^2
\Bigg[
\left(2\,\tilde{c}+1+\frac{\Pi'}{r}\right)\,\mathcal{C}_1
+\frac12\,\left(\tilde{c}^2-1+\frac{4\,\tilde{c}\,\Pi'}{r}+\left(\frac{\Pi'}{r}\right)^2\right)\mathcal{C}_2
\nonumber\\
&\qquad\qquad\qquad\qquad\qquad\qquad\qquad\qquad\qquad\qquad\qquad\qquad+\,\left(\tilde{c}^2-1+\frac{(2\,\tilde{c}-1)\Pi'}{r}\right)\frac{\mathcal{C}_3\Pi'}{r}
\Bigg]\,.
\label{eq:mastereq-reduced}
 \end{align}

\bibliography{nonlinearGMGbib}

\end{document}